\documentclass[12pt]{article}
\usepackage{latexsym,amsmath,amssymb,amscd,cite}
\usepackage{epsf}
\usepackage[all]{xy}

\usepackage{hyperref}

\newcommand{\VolumeHeader}{}
\newcommand{\VolumeSerial}{LNS}

\newcommand{\ActivityName}{ {\normalsize {\it 
Spring School on Superstring Theory and Related Topics  
}}\\
}
\newcommand{\ActivityDate}{ {\normalsize {\it
Trieste, March 31- April 8,2003
}}}

\newcommand{\beq}{\begin{equation}}
\newcommand{\eeq}{\end{equation}}
\newcommand{\bea}{\begin{eqnarray}}
\newcommand{\eea}{\end{eqnarray}}
\def\cM{{\mathcal M}}

\def\cF{{\mathcal F}}
\def\cR{{\mathcal R}}
\def\cL{{\mathcal L}}
\def\cW{{\mathcal W}}
\def\cA{{\mathcal A}}
\newcommand{\cO}{{\cal O}}

\DeclareFontFamily{U}{rsf}{}
\DeclareFontShape{U}{rsf}{m}{n}{
  <5> <6> rsfs5 <7> <8> <9> rsfs7 <10-> rsfs10}{}
\DeclareMathAlphabet\Scr{U}{rsf}{m}{n}
\def\IC{{\mathbb C}}
\def\ZZ{{\mathbb Z}}
\def\IP{{\mathbb P}}
\def\IR{{\mathbb R}}
\def\del{{\partial}}
\def\jb{{\bar j}}\def\l{{\lambda}}
\def\al{\alpha}
\def\be{\beta}

\def\ga#1{\gamma^{(#1)}}
\def\Ga#1{\Gamma^{(#1)}}
\def\gah#1{{\widehat{\gamma}}^{(#1)}}

\def\sst{\scriptstyle}
\def\delh{{\delta_{\hat z}}}
\def\Omthree#1#2#3{{\Omega^{#1,#2}_{{#3}}}}
\def\whx{{\widetilde X}}
\def\bdy{\partial}
\def\xt(#1){\theta_{#1}}
\def\vpi{\omega}
\def\a{\alpha}
\def\xa(#1){{n_{#1}}}
\def\zx{{\xi}}
\def\zo{{z}}
\def\zn{{\hat z}}
\def\xo{{t}}
\def\xs{{\hat t}}

\def\weqn#1{}
\def\eqn#1{}

\newcommand{\foot}{\footnote}

\newcommand{\LectureHeader}{Preparing...}

\begin{document}
\pagestyle{myheadings}
\markboth{\LectureHeader}{\VolumeHeader}
\markright{\VolumeHeader}


\begin{titlepage}


\title{Special Geometry and Mirror Symmetry for Open String 
Backgrounds with N=1 Supersymmetry} 

\author{W.\ Lerche$^\dagger$\thanks{wolfgang.lerche@cern.ch}
\\[1cm]
\\
{\normalsize
{\it $^\ddagger$ Theoretical Physics Division, CERN,
Geneva, Switzerland.}}
\\[10cm]
{\normalsize {\it Lecture given at the: }}
\\
\ActivityName 
\\
\ActivityDate 
\\[1cm]
{\small \VolumeSerial} 
}
\date{}
\maketitle
\thispagestyle{empty}
\end{titlepage}

\baselineskip=14pt
\newpage
\thispagestyle{empty}


\begin{abstract}


We review an approach for computing non-perturbative, exact
superpotentials for Type II strings compactified on Calabi-Yau
manifolds, with extra fluxes and $D$-branes on top. The method is
based on an open string generalization of mirror symmetry, and takes
care of the relevant sphere and disk instanton contributions.  We
formulate a framework based on relative (co)homology that uniformly
treats the flux and brane sectors on a similar footing. However,
one important difference is that the brane induced potentials are
of much larger functional diversity than the flux induced ones,
which have a hidden $N=2$ structure and depend only on the bulk
geometry.

This introductory lecture is meant for
an audience unfamiliar with mirror symmetry. The transparencies
are available at:  
\href{http://wwwth.cern.ch/~lerche/papers}{{\tt http://wwwth.cern.ch/\string~lerche/papers}}.

\end{abstract}

\vspace{6cm}




\newpage
\thispagestyle{empty}
\tableofcontents

\newpage
\setcounter{page}{1}

\section{Introduction}

Most exactly computable quantities in string compactifications have
some underlying ``BPS-property", which leads to special
features of the relevant part of the effective lagrangian. Typically,
these quantities are holomorphic functions\footnote{Rather:
sections, as they are multi-valued.} depending on chiral
superfields, and the rules of chiral superspace integration
imply that such quantities enjoy very special non-renormalization
properties. Either such quantities are not corrected at all at the
quantum level, or only in a very controlled way, for example
perturbative corrections may occur only up to one-loop order. A
closely related feature is that the corrections to these special
terms are only due to BPS saturated intermediate states; these are
under tight analytical control even at the non-perturbative quantum
level, and this is what underlies the exact computability of these
terms of the effective theory.

The prime example of such a holomorphic BPS quantity is the
prepotential $\mathcal F$, which figures as (the two-derivative
part of) the gauge field effective action of $N=2$ supersymmetric
string compactifications, and their various field theory limits.
It gets radiative corrections to one-loop order only, plus an
infinite series of non-perturbative corrections. From the point of
view of the string world-sheet, these non-perturbative corrections
are due to world-sheet instantons, but on the other hand from the
point of view of the effective target space theory, they can often
be interpreted terms of gauge theory instantons.

It is known since ten years or so how to compute exact, non-perturbative
quantum corrections to the $N=2$ prepotential \cite{Candelas:1991rm},
and this has been mainly done by using the methods of mirror symmetry.
As we will briefly review in the following, the main ingredients
that go into this framework are notions such as Calabi-Yau manifolds,
topological field theory, period integrals, the variation of Hodge
structures and "Special Geometry" \cite{deWit:1984pk,Cremmer:1985hj,Ferrara:1989vp,Strominger:1990pd,Bershadsky:1994cx}; for in-depth references on 
mirror symmetry, see for
example
refs.~\cite{mirbook1,mirbook2,Hosono:1994av,Greene:1995yt,Greene:1996cy,Mirbook3}.

More recently there has been dramatic progress in computing analogous
holomorphic quantities also for $N=1$ supersymmetric string and
field theories. The most important such quantity is, of course,
the effective superpotential $\mathcal W$, which determines the
vacuum structure of the theory. Another important holomorphic
quantity is the gauge coupling function $\tau$, which multiplies
the gauge kinetic term: $\cL \sim \tau {\rm Tr} W^\alpha W_\alpha$.
The main new physical ingredient are $D$-branes and background
fluxes, whose presence in a Type II string compactification on a
Calabi-Yau space reduces the supersymmetry from $N=2$ to $N=1$.

Such ``open-closed Type II string backgrounds'' can be crudely
labelled by triples $(X,N_a; \hat N_a)$, where $X$ is some compact
or non-compact Calabi-Yau manifold, $N_a$ denotes the flux numbers
and $\hat N_a$ denotes the brane numbers; the first two entries
pertain to the closed string sector while the last entry pertains
to the open string sector. In addition to these discrete data there
will in general be a number of moduli from both of the closed and
open string sectors.  

These backgrounds may be viewed as specific
parametrizations of an extremely large \cite{Douglas:2003um}
class of four-dimensional
string theories with $N=1$ supersymmetry, and represent toy
models that combine interesting physics, characteristic predictions
and computability of the effective superpotential and gauge couplings.
They are certainly much more complicated as compared to the ordinary
Calabi-Yau backgrounds $(X,0;0)$ with $N=2$ supersymmetry, and this
reflects the growing complexity of theories with less supersymmetry.
This is the reason why it took quite some time until we could
quantitatively access these phenomenologically much more important
$N=1$ supersymmetric theories.  By ``quantitatively'' we mean 
that we now have good analytical control over the vacuum
structure, and the ability to exactly and systematically compute
effects such as non-perturbative supersymmetry breaking\foot{For
an example, see e.g., \cite{Mayr:2000hh}.} for many choices of 
$(X,N_a; \hat N_a)$.

There are by now several techniques available for computing exact
non-perturbative effective superpotentials
 (and other holomorphic quantities related to higher genus world-sheets)
in the presence of $D$-branes,\foot{There are many works dealing
with compactifications of $M$ and $F$ theory, as well brane
constructions, but we will discuss here only Type II strings on
Calabi-Yau manifolds with $D$-branes on top.} which
are interrelated and each of which has certain merits and limitations.
For example, the approaches based on boundary superconformal field
theory \cite{Brunner:1999jq} and on the mathematics of coherent
sheaves \cite{Douglas:2002fr} have provided important insights into
the non-perturbative regime of $D$-brane configurations.  The
approach via large $N$ transitions
\cite{Gopakumar:1998ki,Gopakumar:1998jq,Vafa:2000wi} and Chern-Simons
theory \cite{Ooguri:1999bv,Diaconescu:2002sf,Aganagic:2002qg,Iqbal:2003ix}
has proven to be extremely powerful for a certain class of Calabi-Yau
string compactifications with branes, while matrix models have been
especially useful for describing $N=1$ gauge theories
\cite{Dijkgraaf:2002fc,Dijkgraaf:2002vw,Dijkgraaf:2002dh}.  Very
recently, a powerful framework was developed \cite{AKMV1,ADKMV},
which unifies these and other viewpoints, and this gives an entirely
new way \cite{Iqbal:2003ds} to access the strong coupling regime
of $N=1$ supersymmetric theories.

Less widely known among physicists but powerful as well
is a mathematically inspired method based on the localization
of the path integral on certain fixed points of torus actions;
for some expositions, see e.g. \cite{Graber:2001dw,Mayr:2002zi} (for a
different perspective on localization, see: 
\cite{Nekrasov:2002qd,Eguchi:2003sj}).
Moreover there are geometrical approaches involving mirror symmetry
\cite{Kachru:2000ih,Kachru:2000an,Aganagic:2000gs,Aganagic:2001nx}, 
and in particular there
is also an approach based on a direct generalization of
closed string mirror symmetry to open string backgrounds
\cite{Mayr:2001xk,Lerche:2001cw,Lerche:2002ck,Lerche:2002yw} -
it is this latter approach what we will review in
the present lectures.

Before ending these introductory remarks, we like to add that it is
of course not just computing superpotentials what we may want to do with
these techniques. A conceptually more interesting use is
to get a better understanding of the {\it stringy quantum geometry}
of $D$-branes; that is, the way classical geometrical
notions such as manifold, curvature and gauge field configurations,
are modified due to non-perturbative quantum corrections. Sometimes
such corrections can completely blur out classical notions such
as volume and dimension.

More specifically, it is important to realize that the notions of
classical geometry, such as a $D$-brane wrapping a $p$-dimensional
cycle in a manifold, apply generically only to situations with weak
string coupling, small curvatures and large volumes. If we move
away from the large volume limit of the compactification manifold,
the effects of world-sheet instantons become stronger and stronger
until the classical picture is lost. Similar to what happens in
$N=2$ Yang-Mills theory where we find massless monopoles at strong
coupling \cite{Seiberg:1994rs}, novel phenomena can become visible
in the non-classical regime, which may be better understood by
switching to a more suitable description of them in terms of new,
weakly coupled physical degrees of freedom.  If we move to weak
coupling again, we may end up, due to monodromy, with a completely
``different'' classical brane configuration that wraps a collection
of cycles with dimensions different to the ones we had before; this
shows that in stringy geometry there is no absolute and invariant
notion of the dimension of a $p$-cycle.

Note however, that there is an important difference as compared to
the well-understood theories with $N=2$ supersymmetry; for these,
there is a continuous manifold of vacuum states, the moduli space,
and it makes good sense to consider continuous deformations and
perform analytic continuation into regimes with strong coupling.
On the other hand, for the $N=1$ theories the very presence of a
superpotential represents an {\it obstruction} to continuous
deformations, and this is often a direct consequence of mathematical
obstruction properties of $p$-cycles. Thus it is a priori not clear
to what extent arguments that rely on continuous deformations can
be trusted - even more so because the presence of a superpotential
renders the theory off-shell (unless we happen to sit at a critical
point), for which instance the usual formulation of string theory
based on conformal field theory is not well suited. However, we
will not consider ordinary strings but rather topological strings,
for which being off-shell is not so problematic and for which the
computation of correlators just boils down to certain computations
in algebraic geometry - namely to more or less exactly the ones
that expose the mathematical obstruction properties of $p$-cycles.
One expects that the topological strings are equivalent to the
untwisted ones as far as the BPS sector (e.g., the holomorphic
superpotential in the effective action) is concerned, and that's
anyway all what we expect to be able to compute exactly.
\goodbreak

\section{Recapitulation of $N=2$ ``closed string'' mirror symmetry}

Since the approach for computing $N=1$ superpotentials we want to
explain is an open string generalization of the well-known approach
for computing $N=2$ prepotentials via mirror symmetry, we first
need to review the latter. However, because this is a subject that
is well-understood since quite some time and about which there
exists a collection of excellent review papers and books
\cite{mirbook1,mirbook2,Hosono:1994av,Greene:1995yt,Greene:1996cy,Mirbook3}
(and especially
also because topological field theories were lectured upon in great
detail in Hiroshi Ooguri's lectures in this school), we will be brief
and just quickly go over the key points, with emphasis on the topics
that we will need to know later.

\subsection{Type II strings on Calabi-Yau manifolds}

A Type II string compactification to four dimensions with $N=2$
space-time supersymmetry is described at large radius by a
two-dimensional sigma model with $c=9$ and $(2,2)$ world-sheet
supersymmetry, with target space $X$ (this is the internal, compact
part of the theory only; of course, we implicitly need to add the
non-compact space-time sector plus the appropriate ghost fields).
In order to preserve $N=2$ space-time supersymmetry, the target
space $X$ must be a three complex-dimensional Calabi-Yau manifold,
which more or less by definition has the special property that it
preserves a covariantly constant spinor which can serve as the
supercharge. More precisely, a Calabi-Yau manifold has the following
equivalent properties:
$$
{\rm Calabi-Yau\ manifold}\ \longleftrightarrow 
\begin{cases}
c_1(R)=0\\{\rm holonomy\ group}\, = SU(3)\\
\exists\ {\rm unique\ global\ holomorphic\ form\ }\Omega^{3,0}\in H^{3,0}(X)\\
\end{cases}
$$
besides being a K\"ahler manifold. The latter means that the metric
can be written as derivative of a generating function, the K\"ahler
potential, $g_{i\bar j}=\partial_i\bar\partial_j K$.  Associated
with the metric is the K\"ahler form,  $J^{1,1}=ig_{i\bar
j}dz^id\bar z^{\bar j}\in H^{1,1}(X)$, which as we will see, plays an important
role in the deformation theory of Calabi-Yau spaces.

The $N=2$ effective action in four dimensions contains various
massless fields, notably hyper- and vector-supermultiplets (apart
from fields of the gravitational sector that we will neglect).
These fields correspond to the deformation parameters (moduli) of
$X$, which are one-to-one to certain differential $p,q$-forms
on $X$: 
\beq
\label{eq:difforms}
\omega^{p,q}\ \equiv\ \omega_{i_1,\dots,i_p,\bar j_1,\dots,\bar j_q}dz^{i_1}\wedge\dots
dz^{i_p}\wedge d\bar z^{\bar j_1}\wedge\dots d\bar z^{\bar j_q}\ .
\eeq
Here, $p$ and $q$ count the holomorphic and anti-holomorphic
tensor degrees with respect to the Dolbeault operator $\bar\partial$.

The massless fields in four dimensions correspond to the zero modes
of the laplacian, \def\lb{{\bar\partial}} $\Delta_\lb\ =\
\lb\lb^\dagger+\lb^\dagger\lb$, and thus are given by the closed
but not exact differential forms, ie., by the cohomology of $X$:
$$
H^{p,q}_{\bar\partial}(X,{\IC})\ \equiv\ {
\{\omega^{p,q}|\bar\partial \omega^{p,q}=0\} \over
\{\eta^{p,q}|\eta^{p,q}=\bar\partial \rho^{p,q-1}\}}\ .
$$
On a Calabi-Yau threefold $X$, there are two fundamentally different
kinds of such moduli, namely the
\begin{itemize}
\item K\"ahler moduli (size parameters):  $t_i\sim \omega^{1,1}_i\in H^{1,1}(X)$, $i=1,\dots,h^{1,1}$,
\item Complex structure moduli (shape parameters):  $z_a\sim \omega^{2,1}_a\in H^{2,1}(X)$, $a=1,\dots,h^{2,1}$.
\end{itemize}
The integers $h^{p,q}\equiv {\rm dim}H^{p,q}(X)$ are topological
invariants of $X$ and are called Hodge numbers.

In order to match these two types of deformation parameters to the
hyper- and vector-supermultiplets in the effective lagrangian, we
have to specify about which particular Type II string
compactification we actually talk about. This issue is intimately
tied to the notion of {\it mirror symmetry}, which can be phrased
in a simple way by saying that for ``every''\footnote{While this
seems to be generically true, this statement can be subtle in special
cases, for example for rigid Calabi-Yau's for which there is no
obvious mirror. For such cases the notion of a mirror manifold must
be appropriately generalized.}  Calabi-Yau space $X$, there exists
a mirror partner $\whx$ for which the K\"ahler and complex
structure sectors are exchanged; ie.,
\begin{eqnarray}
H^{1,1}(X)\ &\cong&\ H^{2,1}(\whx)\nonumber\\
H^{2,1}(X)\ &\cong&\ H^{1,1}(\whx)\ .
\end{eqnarray}
The physical meaning of the mirror relation is that the {\it Type
IIA string compactified on $X$ is indistinguishable of the Type
IIB string compactified on the mirror, $\whx$}. In perturbation
theory this statement boils down to a simple sign flip of one of
the left- and right-moving $U(1)$ currents of the 2d $N=(2,2)$
superconformal algebra. However mirror symmetry is a
stronger statement than that, in that it is supposed to hold also
at the non-perturbative level.

An important feature is that K\"ahler and complex structure moduli
fields do not mix in the effective lagrangian, at the two-derivative
level that we consider. Accordingly the scalar sector of the low
energy effective field theory is described by a sigma model whose
target space factorizes:
\beq
\label{eq:totalM}
\cM_{{\rm tot}} \ =\ \cM_H \times\,\cM_V\ ,
\eeq
where $\cM_{H,V}$ denote the spaces of the VEV's of the hyper(H)-
and vector(V) multiplets, respectively. This vacuum manifold can then be matched
to the K\"ahler and complex structure moduli spaces $\cM_{K,CS}$
as follows:\foot{The star on $\cM_H$ means that the
dilaton field, which is a non-geometric "universal" extra
hyper-multiplet not related in any intrinsic way to the Calabi-Yau
manifold, is omitted in the table.}
\begin{center}
\begin{tabular}{cccc}\\
 & Type IIA/$X$ & $\longleftrightarrow$ & Type IIB/$\whx$ \\  
 \hline
 $\cM_H^*:$ & $\cM_{CS}(X)$ & $=$ & $\cM_{K}(\whx)$\\
 $\cM_V:$ & $\cM_{K}(X)$ & $=$ & $\cM_{CS}(\whx)$\\
\end{tabular}
\end{center}
In the following, we will consider only the manifold of the
vector-multiplet VEV's, ie., the lower row of this table.
It is only this sector of the theory that
can be solved for, because the dilaton field belongs to the
hyper-multiplets, and by the above-mentioned factorization theorem
it cannot mix with the vector multiplets. Since its exponential
corresponds to the string coupling, this means that there are no
quantum corrections in the space-time sense to the vector-multiplet
sector of the theory.

However, there are in general world-sheet instanton corrections,
due to 1+1 dimensional string world-sheets that wrap the various
2-cycles $\ga2\in H_2(X)$ of the Calabi-Yau threefold.
The volumes of these cycles are measured by
the K\"ahler parameters $t_i$, $i=1,...,h^{1,1}$, via\foot{More precisely, the $t_i$ should be considered as a complex variables whose additional
imaginary part is provided by the integral over the $B$-field; this will
always be implicitly assumed in the following.}
\beq
\label{eq:spherevol}
t_i\ =\ \int_{\ga2_i}J^{1,1}\ .
\eeq
The instantons thus lead to classical sectors in the path integral weighted by
\beq
\label{eq:sphereinst}
e^{-S_{sphere inst}^{(i)}} \ \sim\ 
e^{-{\rm Volume}(\ga2_i)}\ =\ e^{-t_i}\equiv q_i\ .
\eeq
Mathematically, the world-sheet instantons correspond to maps from
the string world-sheet into the Calabi-Yau manifold $X$, and thus
determining the instanton corrections amounts to count all those
maps in an appropriate way - which is a priori a quite non-trivial
problem.\footnote{Note that the string world sheets can have any
number of holes. We will restrict ourselves here to genus zero
(spherical) world-sheets, which correspond to instanton corrections
to string tree level amplitudes.}

This is where mirror symmetry comes to the rescue, 
as it implies among other things that
\beq
\label{eq:cycleMirr}
H_{even}(X)\ \cong\ H_{3}(\whx)\ .
\eeq
This in particular means that the 2-cycles, around which the Type
IIA world-sheet instantons wrap, map into 3-cycles in the mirror
Calabi-Yau; however, since there are no 2-branes in the Type IIB
string whose 2+1 dimensional world-sheets could possibly wrap those
3-cycles, there cannot be any instanton corrections on the
mirror Type IIB side and hence, the classical computation must be
exact.

In order to be more concrete, we need to specify what physical
quantities we actually talk about. Some of the most basic and
important objects are appropriate integrals $\Pi^\al$ over the
relevant even (Type IIA) or three (Type IIB) dimensional homology
cycles, which physically may be viewed as ``quantum volumes'' and
which mathematically are known as {\it period integrals}. Mirror
symmetry implies that the following two
kinds of integrals must represent the same physical quantities:\foot{
$a,i=1,...,h^{1,1}(X)=h^{2,1}(\whx)$, $\al=1,...,2h^{2,1}(\whx)+2$,
$k=0,...,3$}
\begin{eqnarray}
\label{eq:periodInt}
{\rm IIA/X}:\ \ &\longleftrightarrow  &\ \ {\rm IIB}/\whx:\nonumber\\
\ \Pi^\al\ =\ \int_{{\ga{2k}_i}} (\wedge J^{1,1})^k+{\rm inst.\ corr}\  
&=&
\ \int_{{\ga{3}_\al}} \Omega^{3,0} \\
\sim (t_i)^k +{\mathcal O}(e^{-t}) &  &
\sim (\ln z_a)^k +{\mathcal O}(z). \nonumber\\\nonumber
\end{eqnarray}
The key point is that the quantum volumes of the even dimensional
cycles of $X$ are corrected by world-sheet instantons, while their
images on the mirror side (given by integrals over 3-cycles of $\whx$) are
not corrected. Equating both sides thus allows to determine the
instanton corrections exactly.

The identity (\ref{eq:periodInt}) contains as special case ($k=1$)
the map between the K\"ahler moduli $t_i$ of $X$ and the complex
structure moduli $z_a$ of the mirror $\whx$, commonly called the
``mirror map'':
\beq
\label{eq:mirrormap}
t_i\ =\ -\ln z_a + {\mathcal O}(z).
\eeq

The period integrals, viewed as quantum corrected volume integrals, are
important physical objects for a variety of reasons, and their
detailed study has provided a lot of physical insight. In particular,
they play a dual r\^ole from the perspective of BPS configurations:
at the one hand, they give the exact BPS masses $m_\al\sim |\Pi^\al|$
of $p$-branes wrapped around homology $p$-cycles $\ga p$.
On the other hand, as mentioned above, they determine the effect
of BPS world-sheet instantons corresponding to $(1+1)$-dimensional
world-sheets wrapped around $2$-cycles.

Moreover, the periods form the building blocks out of which the
holomorphic $N=2$ prepotential $\cF$ can be constructed (later we
will see how also the $N=1$ superpotential can be written in terms
of period integrals and generalizations of them). In order to do
so, one first of all needs to pick an integral basis of the homology
3-cycles and group them into two mutually intersecting sets,
$\{\gamma_\al^3\}\ \rightarrow\ \{\gamma_A^3,\gamma_B^3\}$,
$A,B=1,...,h^{2,1}(\whx)+1$. That
is, one chooses a basis of $A$- and $B$-cycles such that their
intersection matrix takes the form
$
\Sigma = \begin{pmatrix}
\ 0 &\ 1 \\
-1 & \hfill\ 0 
\end{pmatrix}\ , 
$
which reflects the symplectic structure\foot{This
carries over to the periods and one may say that the periods are
sections of an $Sp(2h^{2,1}+2,\ZZ)$ bundle over the moduli space.
This means that when varying the moduli, there will be an
$Sp(2h^{2,1}+2,\ZZ)$-valued monodromy action on the periods so that they
are defined only up to such transformations. This ambiguity is a
generalization of $T$-duality.} of the 3-homology. Accordingly, one can write
$$
\Pi^\al(z)\ =\ 
\Big(
X_A,\,
{\mathcal F}^B
\Big)
\ \equiv \
\Big(
\int_{{\ga{3}_A}} \Omega^{3,0},\,
\int_{{\ga{3}_B}} \Omega^{3,0} \Big)(z)\ ,
$$
and in terms of these ``electric'' and
``magnetic'' type of periods the prepotential takes the following
simple form:
\beq
\label{eq:Fz}
{\mathcal F}(z)\ =\ {1\over2}\,X_A {\mathcal F}^A(z)\ .
\eeq
This expression is exact, as there are no instantons of the right kind
to possibly correct this formula.

The more interesting part of the story is when we map back to the
Type IIA side, and all what is necessary for doing so is to determine
the mirror map (\ref{eq:mirrormap}), invert it and substitute
$z=z(t)$ in (\ref{eq:Fz}). This then gives the prepotential in terms
of the Type IIA K\"ahler moduli of $X$, including all the quantum
corrections due to (genus zero) world-sheet instantons. It has the
following general form ($r=h^{1,1}(X)$):
\beq
\label{eq:Ft}
{\mathcal F}(t)\ =\ 
{1\over3!}c^0_{ijk}t_it_jt_k +
\sum_{n_1... n_r} d_{n_1... n_r} 
{\rm Li}_3({q_1}^{n_1}...{q_r}^{n_r})\ .
\eeq
Here the first term is the classical contribution, which is given
by the triple intersections of (dual) 2-cycles in $X$ (actually there are
further lower order terms in the $t$'s that we suppress). The second term
is more interesting, in that the integers $d_{n_1... n_r}$ describe
the contributions of the various world-sheet instantons, ie., they count
maps $\IP^1\rightarrow X$ with multi-degrees $n_1... n_r$.\foot{If the
instantons are not isolated but rather come in continuous families,
then the appropriate interpretation of the $d's$ is in terms of
Euler numbers of the instanton moduli spaces.} Moreover ${\rm
Li}_s(q)\equiv \sum_k k^{-s}q^k$ is the polylogarithm function which
universally takes a certain multi-covering (many-to-one instanton maps) 
into account.
Note that the $d's$ obtained in this way are predictions
for highly non-trivial quantities, and indeed
the mathematicians were subsequently able to verify them for a
large class of Calabi-Yau manifolds, $X$.

\subsection{The Special Geometry of the N=2 vector-multiplet moduli space}

As is well-known, the prepotential $\cF$ can be understood from
three interrelated viewpoints: namely as 

\begin{itemize} 
\item{\bf A) $d=4$ $N=2$ space-time effective lagrangian} 
\cite{deWit:1984pk,Cremmer:1985hj} of the
vector-supermultiplets.  It gives rise to $U(1)$ gauge couplings:
$\tau_{ij}(t)=\del_i\del_j\cF(t)$, to ``Yukawa'' couplings:\foot{This
is loosely speaking - for the $N=2$ compactifications there are no
Yukawa couplings, but rather the $C_{ijk}$ describe certain magnetic
couplings. The traditionally used term ``Yukawa coupling'' refers
to the fact that the $C_{ijk}$ become honest Yukawa couplings when
the Calabi-Yau in question is used for an $N=1$ compactification
of the heterotic string.} $C_{ijk}(t)=\del_i\del_j\del_k\cF(t)$,
and to the K\"ahler potential: $K(t, \bar t)=-\ln[\bar X_A\cF^A-X_A\bar
\cF^A]$.

\item {\bf B)} Generating function of tree-level {\bf topological
field theory (TFT) correlators}:
\beq
\label{eq:cderivF}
\langle O_i O_j O_k\rangle(t)\ =\ \del_i\del_j\del_k\cF(t)\ \equiv
C_{ijk}(t)\ .
\eeq
These three-point functions are essentially the same
(up to raising the index with the constant topological metric: $C_{ijk}=\eta_{kl}{C_{ij}}^l$) as the OPE coefficients, or structure
constants of the {\it chiral ring} \cite{LVW}:
\beq
\label{chRing}
{\cR}:\qquad O_i\cdot O_j\ =\ \sum_k  {C_{ij}}^k(t)O_k\ .
\eeq
The elements $O_i$ of the chiral ring
are nothing but the operators
of the 2d world-sheet theory which are simultaneously primary and
chiral: $G^+_{-1/2}O_i|0\rangle_{{\rm NS}} = G^\pm_{1/2}O_i|0\rangle_{{\rm
NS}} =0$. Their space-time significance is, when we talk about a
nonlinear sigma model whose target space is the Calabi-Yau manifold
$X$, that they correspond to the non-trivial cohomology
elements of $X$ (and thus, partly, to the massless moduli fields
in the effective action); this is because the 2d supercharge can
be associated with the $d$ operator acting on differential forms
on $X$.

More precisely, we need to specify here whether we talk about Type
IIA strings on $X$ or about Type IIB strings on the mirror, $\widehat
X$, and how the different combinations of chiral and anti-chiral
rings that one can assemble out of the left- and right-moving sectors,
map to the respective K\"ahler and complex structure moduli. This
requires to refine the de Rham cohomology (of $d$) into the
Dolbeault cohomology (of $\bar\del$), for which there is the added
notion of holomorphicity. Specifically, a differential form (\ref{eq:difforms})
of (anti-holomorphic,holomorphic) degrees equal to $p,q$ will
correspond to a chiral ring element with (left-moving, right-moving)
$U(1)$ charges given by $p,q$. In terms of the left- and right-moving fermions
$\l$ and $\psi$, of the non-linear sigma model on $X$, an explicit expression
can easily be obtained from (\ref{eq:difforms}) by substituting 
$dz^i\rightarrow\lambda^i$, $d\bar z^{\bar j}\rightarrow\psi^{\bar j}$, plus
furthermore one has: ${d\over dz^i}\rightarrow \lambda_i\equiv
g_{i\bar j}\lambda^{\bar j}$.

If we talk about the Type IIA string, then one considers a topological
twist \cite{Witten:1991zz} of ``type A'' of the 2d $N=(2,2)$
non-linear sigma-model, to the effect that the operators that remain
non-trivial BRST classes correspond to the even-dimensional cohomology,
$H^{i,i}(X)$; the complex structure type operators become BRST
trivial and decouple (at tree level, that is).  Thus, the chiral
ring is of (chiral,chiral)-type and is generated by
$$
O_{A,i}^{1,1}\ =\ \omega^{1,1(i)}_{k\jb}\,
\l^k\psi^\jb\ \in\ H^{1,1}(X)\ ,
$$
which correspond to the K\"ahler deformations. The OP algebra looks:
$$
{\mathcal R}^{(c,c)}:\ \ O_{A,i}^{p,p}\cdot O_{A,j}^{q,q}\ =\ \sum_k C_{ij}^{A\ k}\,O_{A,k}^{p+q,p+q}\ ,
$$
which at first sight one would tend to identify with the cohomology ring
$$
H^{i,i}:\ \ \ H^{p,p}(X)\cup H^{q,q}(X)\ \rightarrow\  
H^{p+q,p+q}(X)\ ,
$$
where the cup product is defined via wedging of forms.  However,
while the charge (degree) structure of these two rings is the same,
the numerical values of the structure constants do in general not coincide,
the reason being the corrections from the world-sheet instantons. One
may rather say that the $(c,c)$ chiral ring is a {\it quantum
deformation} of the classical cohomology ring: ${\mathcal R}^{(c,c)}\cong
QH^{i,i}(X)$.

On the other hand, if we talk about the Type IIB mirror theory on
$\whx$, then one can implement the ``B type'' topological
twist, to the effect that only the complex structure type of operators
survive as physical operators; the K\"ahler type operators become
BRST trivial and decouple. The chiral ring
\beq
\label{eq:BtypeRing}
{\mathcal R}^{(a,c)}:\ \ O_{B,a}^{-p,p}\cdot O_{B,b}^{-q,q}\ =\ 
\sum_c C_{ab}^{B\ c}\,O_{B,c}^{-p-q,p+q}
\eeq
is of (anti-chiral, chiral)-type\foot{Note that a negative form
degree (charge) can be converted to a positive one by contraction
with the holomorphic 3-form:  $\Omega^{3,0}: \omega^{-p,q}
\rightarrow \omega^{3-p,q}$.} and is generated by:
$$
O_{B,a}^{-1,1}\ =\ {\omega^{-1,1(a)}}^i_{{\jb}}\,\l_i\psi^\jb\, \in \,
H^{-1,1}(\whx)\cong H^{2,1}(\whx)\ ,
$$
which correspond to the complex structure deformations. Due to the
absence of any world-sheet instanton corrections in the $B$-model,
the $(a,c)$ ring OPE coefficients ${c_{ab}}^{B\ c}$ are identical
to the ring structure constants of the classical cohomology ring:
$$
H^{-p,p}(X)\cup H^{-q,q}(X)\ \rightarrow\  H^{-p-q,p+q}(X)\ .
$$

Now recall that, by mirror symmetry, the $A$-model on $X$ is
equivalent to the $B$-model on $\whx$, and thus we can equate
the quantum deformed cohomology ring on $X$ with the classical
cohomology ring on $\whx$, ie.,
$$
{{\mathcal R}^{(c,c)}}(X)  \equiv\ QH^{i,i}_{{\bar \partial}}
(X)\ \cong\ 
H^{-j,j}_{{\bar \partial}}(\whx)\ \equiv\ 
{{\mathcal R}^{(a,c)}}(\whx) .
$$
This exhibits in mathematical terms how we can determine the quantum
corrected prepotential on the Type IIA side: namely by relating its
triple derivatives (the quantum deformed $(c,c)$ ring structure
constants) to the three-point correlators (classical $(a,c)$ ring
structure constants) on the Type IIB side
$$
\del_i\del_j\del_k\cF(t)\ \equiv\ C_{ijk}^{A}(t)\ =\ \sum_{a,b,c}\,
{\partial z_a\over \partial t_i} 
{\partial z_b\over \partial t_j} 
{\partial z_c\over \partial t_k} \
C_{abc}^{B}(z(t))\ .
$$
The latter are given by the following classical integral:
\beq
\label{eq:Cabc}
C_{abc}^{B}(z)\ =\ \int_X\Omega^{3,0}(z)\wedge \del_a\del_b\del_c \,\Omega^{3,0}(z)\ .
\eeq
This is closely related to the third viewpoint.

\item {\bf C)} The viewpoint of {\bf variation of Hodge structures.}

It is instructive to contemplate upon the previous formula - if
there were no derivatives, then the resulting integral would vanish,
because we need to
have a $(3,3)$ volume form to integrate over. The point is that the
missing $(0,3)$-component is generated by taking three derivatives
of $\Omega^{3,0}$.  This is because the very definition of what
constitutes a holomorphic form (or more generally, a $(p,q)$-form)
within the full $H^3$ changes smoothly when we vary the complex
structure moduli. In particular, upon a first order variation, the
holomorphic $(3,0)$ form gains a component in $H^{2,1}$, and so on.
Altogether one can write this "variation of Hodge structures"
schematically as follows:
\begin{eqnarray}
\label{eq:Hodgevar}
\Omega^{3,0}(z) & \in & H^{3,0}\nonumber\\
\delta_z \Omega^{3,0}(z) & \in & H^{3,0}\oplus H^{2,1}
\\
(\delta_z)^2 \Omega^{3,0}(z) & \in & H^{3,0}\oplus H^{2,1}
\oplus H^{1,2}
\nonumber\\
(\delta_z)^3 \Omega^{3,0}(z) & \in & H^{3,0}\oplus H^{2,1}
\oplus H^{1,2}\oplus H^{0,3}\ .
\nonumber
\end{eqnarray}
Note that after three variations there is indeed an anti-holomorphic
$(0,3)$ component generated, and it is precisely this component what then leads to a non-vanishing integral in (\ref{eq:Cabc}).

Note also that the sequence of variations terminates after the
third step, in that the complete third cohomology, $H^3$, has been
generated and no further different $p,q$ forms can be generated
by doing further variations. This means that the fourth derivative
of $\Omega^{3,0}$ must be expressible in terms of the same set
of differential forms, and thus must be expressible in terms of
lower order variations - which just means that $\Omega^{3,0}$
must obey some linear differential equation.

Specifically, fixing an ordered, $2h^{2,1}+2$ dimensional basis 
for $H^3$:
\beq
\label{eq:Racbasis}
\vec\vpi\ =\ \big(\Omega^{3,0}, \Omega^{(2,1)}_a, 
\Omega^{(1,2)\,a},
\Omega^{(0,3)}\big)^t\ ,\qquad a=1,...,h^{2,1}(\whx)\ ,
\eeq
(where $\Omega^{(2,1)}_a$ correspond to the $(a,c)$ ring elements
$O_{B,a}^{(-1,1)}$ in the topological field theory), we can translate
the above pattern of Hodge variations into a matrix differential
equation of the following form:
\beq
\label{eq:matdeq}
\nabla_a\cdot\vec\vpi(z)\ \equiv\ \big[\del_a- \cA_a(z)\big]\cdot
\vec \vpi(z)\ =\ 0\ .
\eeq
This holds only up to exact pieces that can be freely added to
differential forms. However, the exact pieces drop out once we
consider instead integrals over an appropriate
fixed basis of 3-cycles, i.e.:
\beq
\label{eq:periodm}
\Pi^\al_\be(z)\ =\ \int_{\gamma_\al^{(3)}}\vpi_\be(z)\ ,\qquad
\gamma_\al^{(3)}\in H_3(\whx)\ ,\vpi_\be\in H^3(\whx)\ .
\eeq
This ``period matrix'' then satisfies $\nabla_a\cdot \Pi^\al_\be=0$
provided the 3-cycles have no boundaries (which is the case for the
closed string theories at hand, and one of the new ingredients that
we will find later in the case of open string backgrounds with
$D$-branes, is that there will be extra contributions from non-zero
boundaries of 3-cycles).

The nice thing about these differential equations is that
it is known from general mathematical theorems about the variation
of complex structures, that:
\beq
\label{eq:flatness}
\big[\,\nabla_a,\nabla_b\,\big]\ =\ 0\ .
\eeq
Thus the complex moduli space is ``flat'' - meaning that one can
view this equation as a zero curvature, or integrability property
of the matrix system (\ref{eq:matdeq}). From this viewpoint one may
split the covariant derivative into a "Gau\ss -Manin connection"
piece plus a remainder, $C$:
$$
\nabla_a\equiv \del_{z_a}- \cA_a(z)\ =\ \del_{z_a}-\Gamma_a(z)-C_a(z)\ ,
$$
the two pieces being distinguished by the type of non-zero entries they have, i.e.:
$$
\Gamma_a=
\begin{pmatrix}
* &\ &\ &\ \\
* & * &\ &\ \\ 
* &* &* &\ \\
*\ &*\ & *\ & *\
\end{pmatrix} 
\ ,\qquad
(C_a)_\be^{\ \gamma}=
\begin{pmatrix}
\ &1 &\ &\ \\
  &  &(C_a)_b^{\ c} &\ \\ 
  & &\ 1 &\ \\
\phantom{*\ }  & \phantom{*\ } &\phantom{*\ }   & \phantom{*\ }
\end{pmatrix}\ .
$$
Note that the matrices $C_a(z)$, which correspond
to the coefficients of the rightmost terms in the list (\ref{eq:Hodgevar})\
of Hodge variations, are nothing but the cohomology ring structure
constants of $H^3(\whx)$, and physically they are precisely the OPE coefficients
(\ref{eq:BtypeRing}) of the $B$-model chiral ring, $\cR^{(a,c)}(\widehat
X)$.

As a side remark, note also that the differential equations have a
concrete meaning also completely within topological field theory.
An important property of the TFT is the existence of a flat connection
on the bundle of Ramond-Ramond vacuum states, fibered over the
moduli space. From this viewpoint, $[\del_a-C_a(z)]\simeq 0$ expresses
that a derivative with respect to a deformation parameter amounts
to an operator insertion in the path integral, and this insertion
is represented by the ring structure constant matrix, $(C_a)_\be^{\,
\gamma}$. Indeed, $(C_a)_\be^{\, \gamma}$ indeed represents the 
chiral primary field
$O_a$ when acting on a given basis of the chiral ring.  For further
reading about the rich ``$tt^*$ geometry'' of $N=(2,2)$ superconformal
field theories we refer to \cite{Bershadsky:1994cx,Cecotti:1991me}.

Now, the zero curvature condition (\ref{eq:flatness}) implies that there exist
``flat coordinates'' $t_i(z_a)$, in terms of which the connection piece
$\Gamma$ vanishes:
\beq
\label{eq:matrixEq}
\big[\,\del_{t_i} - C_i(t)\,\big]\cdot \Pi^\al_\be(z(t))\ =\ 0\ ,
\eeq
in conjunction with $\big[\del_{t_i},C_j(t)\big]=\big[C_i(t),C_j(t)\big]=0$. 

As the reader might have guessed from the notation, the flat
coordinates $t_i=t_i(z)$ are precisely the K\"ahler parameters of
the $A$-model on the threefold $X$, that we had defined in and above
of (\ref{eq:mirrormap}).  The make this more concrete, note that
due to the triangular structure of the matrix differential equation,
we can recursively solve for the higher rows of the period matrix $\Pi^\al_\be$
in favor of its first row, which is nothing but the vector of quantum
volumes we introduced earlier:  $\Pi^\al_1\equiv
\Pi^\al=\int_{\gamma_\al^{(3)}}\Omega^{3,0}$. By this iterative
procedure the first order matrix system (\ref{eq:matdeq}) turns
into an equivalent system of higher order differential equations
for the period vector:
\beq
\label{eq:PFeqs}
\cL_a(z,\del_z)\cdot \Pi^\al(z)\ =\ 0\ ,
\qquad a=1,...,h^{2,1}(\whx).
\eeq
This system is commonly called ``Picard-Fuchs system''; in order
to determine the periods, or quantum volumes, it turns out that it
is often easier to solve these differential equations rather than
performing complicated multi-dimensional integrals.

It is a general property of the period integrals that $h^{2,1}(\widehat
X)=h^{1,1}(X)$ of them  behave near the limit of large complex
structures (where $z_a\sim 0$) like $\ln z_a$ (c.f., (\ref{eq:mirrormap})).
These are the ones to be associated with the flat coordinates, or
K\"ahler parameters of $X$: $t_i\sim \ln z_a+\cO(z)$.
Actually there is one unique period
(denoted by $X_0$) that is a pure series in $z_a$ without a logarithm, and one can
use this to normalize the period integrals by dividing them by it.
This means that after normalization the unique special
period will turn into a constant, so that the period vector,
when expressed in terms of the flat coordinates, looks schematically
as follows:
\begin{eqnarray}
\label{eq:Periodst}
\Pi^\al(z(t))\ =\ (X_A,\cF^B)\ &=&(X_0,X_a,\cF^b,\cF^0)
\nonumber\\
&\longrightarrow &(1,t_i,\del_i\cF,2\cF-t^j\del_j\cF)
\ \simeq\ (1,t,t^2+\cO(e^{-t}),t^3+\cO(e^{-t}))\ .
\nonumber\\
\end{eqnarray}
This reproduces the prepotential as promised in (\ref{eq:Fz}), ie.,
$\cF(t)={1\over2}X_A\cF^A(z(t))$. Moreover,
 the complete period matrix takes the form:
\beq
\Pi^\al_\be\ \equiv\   \int_{\ga3_\al}\, \vpi_\be\ 
=\ 
\begin{pmatrix}
1&t_a&\cF^a&2\cF -t_a\cF^a\\
0&\delta_{a}^b&\cF^{ab}&\cF^b-t_a \cF^{ab}\\
0&0          &\cF^{abc}   &-t_a \cF^{abc}\\
0&0&...&...\\
\end{pmatrix}\ ,
\eeq
which solves \ref{eq:matrixEq}\ upon substituting $c_{ijk}\rightarrow
\cF^{abc}$. 
\end{itemize}

Now, having reviewed some of the main aspects of $N=2$ Special
Geometry, we are prepared to proceed into new territory by incorporating
fluxes and $D$-branes.

\section{Open/closed string backgrounds with $N=1$ supersymmetry }

So far we considered a closed string background resulting in an
$N=2$ supersymmetric theory in four dimensions, namely Type II
strings on a Calabi-Yau manifold $X$.  We now seek to reduce the
supersymmetries to $N=1$. More precisely, what we mean is that we
seek a  modification of this $N=2$ supersymmetric background that
can be described in terms of an effective $N=1$ supersymmetric
lagrangian with a holomorphic superpotential; this superpotential
may or may not break supersymmetry even further.  In fact there
are two sorts of such modifications, which we will discuss in turn.

\subsection{Adding background fluxes}

The first modification is a deformation in the closed string theory,
obtained by "switching on" certain background fluxes. This means
that we take $\int_{\ga p} H^{(p)}\not=0$, where $H^{(p)}=dC^{(p-1)}$
are the field strengths of the Type II anti-symmetric tensor
fields. Such backgrounds will generically induce non-zero potentials 
\cite{Polchinski:1996sm}
and trigger subsequent supersymmetry breaking. Which fluxes one can 
switch on depends of course on which Type
II string we consider; more precisely, for the Type IIA string we
can have $p=3,7$ in the NSNS- and $p=2,4,6,8$ in the RR-sector, and for
the Type IIB string we can have $p=1,3,5,7$ in the NSNS- and $p=1,2,5,7,9$
in the RR-sector; see
S.~Trivedi's lectures at this school for further details and an extensive discussion.

Let us first consider the Type IIB string compactified on a threefold
$\whx$, and focus on the fluxes associated with the 
2-form gauge fields. It is well known that the RR and NSNS fluxes
fit nicely together in the form $H^{(3)}=H^{(3)}_{RR}+\tau H^{(3)}_{NS}$,
where $\tau=C^{(0)}+ie^{-\varphi}$ is the complexified Type IIB
coupling field ($\varphi$ is the dilaton). In the following we take
$H^{(3)}_{NS}=0$, so that we are left with RR fluxes only which are
quantized.

It is known \cite{Taylor:1999II,Mayr:2000hh} that such fluxes lead to an 
$N=1$ superpotential of the form:
\beq
\label{eq:WBfluxpot}
\cW_{IIB/\whx}(z)\ =\ \int_{{\whx}}
 H_{RR}^{(3)}\wedge  \Omega^{3,0} =\ \sum_\al N_\al\cdot\Pi^\al(z)\ ,
 \qquad N_\al\in\ZZ\ ,
\eeq
which can be non-vanishing only if the $(0,3)$ component of 
$H_{RR}^{(3)}$ is non-zero; general conditions on fluxes are
discussed in \cite{Gukov:1999ya}.
For us, the main point is that the
3-form flux numbers $N_\al$ simply multiply the periods $\Pi^\al(z_a)$ of
the holomorphic $(3,0)$-form $\Omega^{3,0}(\whx)$.

Note that this modification is purely within the closed string
background, and may be viewed as a spontaneous breaking of
$N=2$ to $N=1$ supersymmetry. That is, apart from the choice
of flux numbers, the theory depends only on the intrinsic "bulk" geometrical
data of the Calabi-Yau threefold $\whx$. Moreover, it depends
only on the complex structure deformations $z_a$ from the closed
string sector which enter via  $\Omega^{3,0}(\whx)$, and it specifically
does not depend on the K\"ahler deformations. This decoupling
property is inherited from $N=2$ supersymmetry,
and is why one can make use of mirror symmetry also
for $N=1$ supersymmetric flux compactifications.

The Type IIA picture of the above superpotential is obtained via
mirror symmetry, and given, essentially, by replacing in the above
formula the period by volume integrals, i.e.,
\begin{eqnarray}
\label{eq:Afluxpot}
\cW_{IIA/X}(t) &=& \int_X \sum_{k=0}^3H_{RR}^{(2k)}(\wedge J^{1,1})^{3-k}
+\dots\\
&=& N^{(6)}+N^{(4)}t+N^{(2)}t^2+N^{(0)}t^3 +{\mathcal O}(e^{-t})\ .
\nonumber\end{eqnarray}
Here $N^{(2k)}$ denote the RR-fluxes through the $0,2,4,6$-cycles,
and the dots indicate that the classical integrals get corrected
by contributions from the world-sheet instantons. A priori there
would be little guidance of how to compute these - were it not for mirror
symmetry (\ref{eq:WBfluxpot}), which via (\ref{eq:Periodst}) implies
that the non-perturbative completion of (\ref{eq:Afluxpot}) must
be given by:
$$
\cW_{IIA/X}(t)=\cW_{IIB/\whx}(z(t))=
N^{(6)}+N^{(4)}t+N^{(2)}\del_t\cF(t)+N^{(0)}\cF^0(t)
$$
with $\cF^0\equiv 2\cF-t\cdot \del_t\cF$.

One may wonder how we could dare to make statements like those
above, despite they deal with non-trivial RR backgrounds for which
there is no good description in terms of a world-sheet theory, and
which in addition may involve some intractable back-reaction on the
Calabi-Yau geometry. The point is that the topological field
theory setup, where we focus only on holomorphic quantities at string tree
level, is insensitive to such back-reaction, and there is a simple
argument \cite{Vafa:2000wi} in terms of the effective action that
shows that the superpotential indeed arises as advertised, irrespective
of how the world-sheet theory is precisely defined. For this, let
us for simplicity focus only on the 2-flux and first state that
while the corresponding CFT vertex operator is not given by a
marginal operator, it is rather given by a massive, "unphysical"
vertex operator which corresponds to an auxiliary field. Schematically,
we can write the modulus superfield as follows: $\Phi=t+\theta^2
H_{RR}^{(2)}$. Thus, if the flux is non-zero: $\langle H_{RR}^{(2)}\rangle
= N^{(2)}$, then indeed: $\int d^4\theta \cF(\Phi)\rightarrow\int
d^2\theta N^{(2)} \del_\Phi\cF(\Phi) \equiv\int d^2\theta\cW(t)$.

\subsection{Adding $D$-branes}

The other possibility to reduce the $N=2$ supersymmetry to $N=1$
is to add a $D$-brane background (a ``$D$-manifold''), which
introduces an open string sector into the theory. A $D$-brane
configuration will in general contribute its own moduli (i.e., related
to location and Wilson line and gauge bundle data), on top of the
moduli that are intrinsic to the embedding bulk (Calabi-Yau) geometry.
It is thus an important question to ask about the structure of the
combined open plus closed string parameter space, and how the open
and closed string moduli fields enter in the effective lagrangian.

A $D$-brane background consists
of one or several $D$-branes that wrap $p$-cycles $\ga p\in
H_{p}(X)$, and otherwise span the non-compact 3+1 dimensional space-time;
the world-volume of a single brane will describe a 3+1
dimensional theory with $N=1$ supersymmetries if the cycle is {\it
supersymmetric}.\foot{If we consider a collection of such cycles, wrapped
by several $D$-branes, then the amount of unbroken supersymmetry
depends also on the intersection properties of these cycles; for simplicity, we
will consider only a single brane, mostly, or parallel branes.} This amounts
to the existence of a covariantly constant spinor $\eta$, which can serve as
an unbroken supercharge. The relevant equations boil down to:
$(1-\Gamma)\eta=0$, where $\Gamma={1\over\sqrt h}\epsilon^{\alpha_1...\alpha_{p+1}}
\partial_{\alpha_1}x^{m_1}...\partial_{\alpha_{p+1}}x^{m_{p+1}}
\Gamma_{m_1...m_{p+1}}$. Here, $h$ is the induced metric on the world-volume, the
derivatives of the coordinates $x$ describe a pull-back to the world-volume, and 
$\Gamma_{m_1...m_{p+1}}$ denote the 10d gamma matrices.

It turns out \cite{Ooguri:1996ck}, by analyzing the solutions to this condition,
that there are two kinds of supersymmetric $p$-cycles on a Calabi-Yau
space, and this reflects 
the decoupled K\"ahler and complex structure sectors of the moduli space.
Correspondingly, $D$-branes wrapped on such cycles are called $A$-
or $B$-type of $D$-branes: 

\begin{itemize} \item{\bf A-type branes}
wrap {\it special lagrangian (SL) cycles}. These are generally
middle dimensional, which means three dimensional for Calabi-Yau
threefolds: $\ga3\in H_3(X)$. There are some extra conditions,
like that the K\"ahler and holomorphic three-forms vanish when
pulled back to the cycle, but this won't be important for our discussion.
As far as the moduli of the brane configuration go, we know a priori
that there should be dim$_{\IR}(\cM_{\ga3})=b_1(\cM_{\ga3})\equiv\hat r$
deformations of a SL cycle. This could be odd, but certainly we
need complex moduli fields in a supersymmetric theory, so at first sight
this seems to pose a problem. The resolution
is that these moduli pair up with extra Wilson line moduli, which
come from a flat $U(1)$ gauge field on the world-volume. In total
we get a doubling of real scalar fields to yield complex ones, 
and we denote these by $\hat t_k$, $k=1,...,\hat r$.
The hat signifies the open string sector, and the letter $t$ reflects
that these moduli are the analogs of the K\"ahler moduli of the
Calabi-Yau space.

\item{\bf B-type branes} wrap {\it holomorphic cycles}, which can
be 0,2,4,6-dimensional on a Calabi-Yau threefold. Besides the
holomorphic embedding geometry, there appears more structure - as
is well-known, for $n$ branes there is a $U(n)$ gauge symmetry
from the open string sector, and the topological sectors of the
gauge field configurations are extra important data of the brane
configuration. In fact, due to the anomalous WZ couplings $\int
C\wedge Tr[e^F]\wedge \sqrt{\hat A(R)}$ on the world-volume of
$B$-type of branes, such gauge field configurations (instantons or
some more general, non-trivial gauge bundles) of a $p$-brane
correspond to bound states of this $D$-brane with lower dimensional
ones; however, delving into this interesting issue would lead us
too far away, and we refer the reader to the literature 
\cite{Douglas:1995bn,Witten:1996im}.  \end{itemize}

We should be careful to add that notions such as ``branes wrapping $p$-cycles
and gauge configurations on top of them'' make sense only in the
semi-classical regime, where curvatures are small and the string
coupling small. Away from this regime there will
be quantum corrections that can wipe out the original meaning of these
notions, and one should adopt a better-suited language of quantum
geometry that takes over in the strong coupling regime. As we will see, a useful
tool for accessing this regime is provided by the open string analog
of mirror symmetry.

For the time being, let us consider only branes with trivial bundle
structure, and moreover restrict to a class of certain specific
$D$-brane geometries.  These geometries were introduced first in
refs.~\cite{Aganagic:2000gs,Aganagic:2001nx}
 and are given by certain non-compact $D$-branes on
non-compact Calabi-Yau manifolds.  As these are best understood
in terms of toric geometry, which is a broad field outside the scope
of these lectures, we refrain from explaining them in more depth
but rather refer the interested reader to these references for more
details. For our purposes suffice it to summarize a few key features
of the $D$-brane setup.

For the $A$-type of branes we consider $D6$ branes (which have a
6+1 dimensional world-volume) that wrap special lagrangian 3-cycles
of a threefold $X$ and otherwise stretch over the remaining 3+1
dimensional space-time. We parametrize the relevant open string
deformations by the sizes of disks $D^{(2)}=\gah2$ whose boundaries lie on
the given 3-cycle:
\beq
\label{eq:diskint}
\hat t_k\ =\ \int_{\gah2_k} J^{1,1}\ .
\eeq
This is entirely analogous to the definition of the closed string,
``bulk'' K\"ahler parameters $t_i$ defined in (\ref{eq:spherevol}),
the only difference being that the open string K\"ahler moduli are
not integrated over holomorphic spheres but over
 holomorphic disks with a boundary (and similarly, as discussed
 above, the moduli get complexified by adding the Wilson line moduli
 as imaginary components). Obviously, when we move a given SL
 $D$-brane within its homology class, the size of the disks attached
 to it will change, which then corresponds to different values of
 the $\hat t_k$.

The open string K\"ahler moduli are analogous to the closed string
ones
 also with respect to world-sheet instantons: if we consider open
string tree-level amplitudes which are given by CFT correlators on
world-sheets with disk topology, then there will be extra
 non-perturbative sectors in the path integral that arise from
 holomorphic maps from the disk-like world-sheets to disks within
 the Calabi-Yau whose boundaries sit on the given SL 3-cycle
 \cite{Kachru:2000ih,Kachru:2000an,Aganagic:2000gs} - exactly the
 same kind of disks that are integrated over in (\ref{eq:diskint}).
 In other words,  there will be open string instanton contributions
to tree-level amplitudes which are weighted by
 \beq
\label{eq:diskinst}
e^{-S_{disk\ inst}^{(k)}} \ \sim\ 
e^{-{\rm Volume}(\gah2_k)}\ =\ e^{-\hat t_k}\equiv \hat q_k\ , 
\eeq
besides the well-known closed string world-sheet instantons with
spherical world-sheets.\foot{And more generally, to string amplitudes
on world-sheets with genus $g$ and $h$ boundaries, there will be
contributions of instantons of the corresponding topology. 
In this lecture we will stick to open string tree level, with $g=0,h=1$.}

Specifically, the disk partition function, which is none other than
the superpotential, has the schematic form:
\beq
\label{eq:superW}
\cW(t,\hat t)\ =\ 0\,t\cdot \hat t+
\sum_{n_i;\hat n_k}d_{n_1,...,n_r;\hat n_1,...,
\hat n_{\hat r}} {\rm Li}_2({q_1}^{n_1}...{q_r}^{n_r}
{\hat q_1}^{\hat n_1}...{\hat q_r}^{\hat n_{\hat r}}).
\eeq
The first term symbolically stands for the classical perturbative
contribution, and the coefficient equal to zero signifies that
classically, there is no superpotential - the reflects a mathematical
theorem that there is no obstruction in deforming a SL 3-cycle.  On
the other hand, the presence of the instanton sum, which represents
a non-perturbative obstruction to deformations, signifies that the
mathematical theorem is violated at the quantum level - this may
be seen as another manifestation of "stringy quantum geometry".

The dilogarithm function takes the multi-coverings (many-to-one
instanton maps) appropriately into account, and when the instanton 
sum is parametrized
in this way, the coefficients $d_{n_1,...,n_r;\hat n_1,...,\hat
n_{\hat r}}$ are integers. Their indices label
``relative'' homology classes of the instanton maps, ie., they
correspond to 2-cycles that are closed only up to boundaries lying
on the SL 3-cycle, $\ga3$. That is, while $n_i\in H_2(X)$ label the usual
homology classes of spheres in $X$, the $\hat n_k\in H_1(\ga3)$
label homology classes of the boundaries of the disks lying on
$\ga3$. To {\it a priori} determine these numbers is a formidable
mathematical problem, and we will show later how they can be
efficiently computed for sufficiently simple geometries via mirror
symmetry.\foot{Very recently, novel methods have been devised 
\cite{Aganagic:2002qg,AKMV1,ADKMV} that
are even more efficient.}

Let is now consider $B$-type branes that wrap holomorphic
cycles. It is known \cite{WittenCS} that for branes wrapping the whole
Calabi-Yau manifold (ie., a 6-cycle), the open string disk partition function
is given by holomorphic Chern-Simons theory:
\beq
\label{eq:ChernSimons}
\cW\ =\ \int_X\Omega^{3,0}\wedge{\rm Tr}\big[A\wedge\bar\del A+
\frac23A\wedge A\wedge A\big]\ ,
\eeq
where $A$ is the gauge field on the world-volume of the $D6$-brane
(or a stack of several $D6$-branes). For lower dimensional branes
wrapping holomorphic submanifolds, one needs to appropriately
dimensionally reduce this expression, and this just amounts
to replacing the relevant components of $A$ by scalar fields $\phi$. We
will specifically consider only $B$-type of branes wrapped around
holomorphic 2-cycles $\ga2$, for which dimensional reduction gives 
\cite{Aganagic:2000gs}
$$
\cW\ =\ \int_{\ga2(\hat z)}\ \Omega^{3,0}_{ijz}\phi^i\bar\del_z\phi^jdzd\bar z\ .
$$
This can be locally rewritten as
\beq
\label{eq:chainint}
\cW(z,\hat z)\ =\ \int_{{\gah3(\hat z)\atop\del \gah3=\ga2}}\Omega^{3,0}(z)\ ,
\eeq
where $\gah3(\hat z)$ is a 3-chain, whose boundary consists of the
2-cycle $\ga2$ around which we wrap the brane. Moreover, $\hat z$
governs the size of the 3-chain, by
parametrizing the location of $\ga2$ within its homology
class.\foot{Actually the situation is a bit more involved
than that, and strictly speaking one needs to include a fixed
reference 2-cycle $\gamma^{(2)*}$ and consider only the difference of $\cW$
between the 2-cycles. We refer the reader to 
\cite{Witten:1998jd,Aganagic:2000gs} for details.} An important
point to note is that (\ref{eq:chainint})\ is quite similar to the
usual period integral (\ref{eq:periodInt}), the only difference
being that it is not a closed 3-cycle over which we integrate but
rather a 3-chain, ie, a 3-cycle with boundary. And analogously,
upon analyzing zero mode structures, one finds that there are no
possible corrections of world-sheet instantons to (\ref{eq:chainint}),
so that it is an exact result - recall however, that despite the
similarity, there is a important physical difference, in that now
we talk about a quantity of an $N=1$ supersymmetric theory rather
than about one with $N=2$ supersymmetry.

In summary, we have seen that the extra K\"ahler- and complex
structure-like moduli $\hat t,\hat z$ coming from the $D$-brane
geometry are quite similar to the closed string, ``bulk'' moduli
$t,z$ -- this suggests that mirror symmetry might be similarly
successfully applied as before to the closed string.

\subsection{Mirror symmetry of $D$-brane configurations}

Recall that in the closed string, ``bulk'' sector, an important feature as to why
mirror symmetry was at all useful, was the fact that the K\"ahler and
the complex structure sectors are decoupled (up to the two-derivative level
of the effective lagrangian). That is, the total moduli space was a direct
product: $\cM_{{\rm tot}}=\cM_{{\rm KS}}(t)\times \cM_{{\rm CS}}(z)$.

In order to successfully apply mirror symmetry to the open string sector as well,
we need to have analogous properties to hold for the open string
K\"ahler and the complex structure type moduli, $\hat t, \hat z$. In fact it can
be argued \cite{Brunner:1999jq}
 with methods of boundary CFT that the requisite decoupling properties
indeed hold (at least at tree level). Specifically, the dependence
of D- and F-terms in the effective lagrangian on K\"ahler and complex structure
type of moduli is as follows:
\begin{eqnarray}
\label{eq:decoupling}
&A{\rm -branes}:\ \
&\begin{cases}
\cW(t,\hat t)&{\rm holom.\ F-term\ potential}\\
D(z,z^*,\hat z,\hat z^*)&{\rm Fayet-Iliopoulos\ D-term}\\
\end{cases}
\\
&B{\rm -branes}:\ \
&\begin{cases}
\cW(z,\hat z)&{\rm holom.\ F-term\ potential}\\
D(t,t^*,\hat t,\hat t^*)&{\rm Fayet-Iliopoulos\ D-term}\\
\end{cases}\nonumber
\end{eqnarray}
From this we see that there is indeed a chance to exactly determine
the instanton corrections to the superpotential 
(and also to other holomorphic quantities), by using
mirror symmetry for mapping $\cW(t,\hat t)$ on the Type IIA string side
to the non-corrected superpotential $\cW(z,\hat z)$ on the Type IIB side.

\begin{figure}[htb]
\epsfysize=2.4in
\centerline{\epsffile{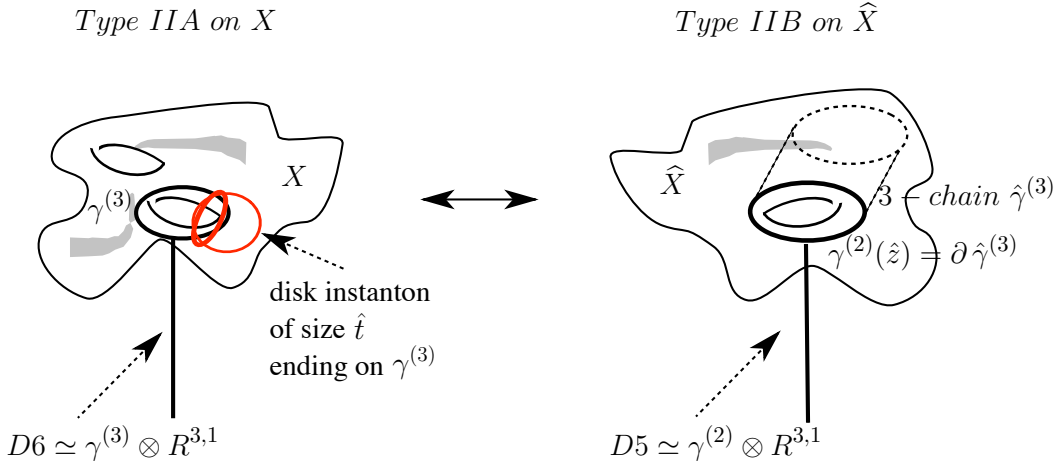}}
\caption{\footnotesize Sketch of the mirror pair of the $D$-brane configurations
we consider.  In the $A$-model on the left side, disk and $S^1$
instantons deform the quantum geometry of the SL 3-cycle $\ga3$,
around which a $D6$-brane is partially wrapped.  On the other hand,
for the $B$-model on the right side the brane geometry remains
uncorrected. Note that this is a simplified picture, in that for
the concrete physical models under consideration, the Calabi-Yau
manifolds and the relevant cycles are non-compact. \label{fig:FIG1}}
 \end{figure}

The important question is how mirror symmetry acts on the
open/closed string background in the
Type IIA string picture.  In fact it so happens for the specific
geometries we consider, that the Type IIB mirror of the SL 3-cycle
of the Type IIA picture is a holomorphic 2-cycle of the kind we
considered in the previous section; see Fig.\ref{fig:FIG1} for a visualization
of the geometry.  Thus we can impose mirror symmetry by requiring
the two expressions
(\ref{eq:superW}) and (\ref{eq:chainint}) for the superpotential
to be equal:
\beq
\label{eq:Wmirror}
\cW_{IIA/X,D6}(t,\hat t)\ \equiv\ 
\sum_{n,\hat n}d_{n,\hat n}{\rm Li}_2[q^n\hat q^{\hat n}]\ \  \mathop{=}^{{!}}\
\int_{{\gah3(\hat z)}}\Omega^{3,0}(z)\ \equiv\
\cW_{IIB/\whx,D5}(z,\hat z)\ .
\eeq
As discussed before, the LHS is corrected by sphere and disk
instantons, while there are no instantons that could possibly correct
the RHS. Thus we have managed to represent the exact $N=1$ $D$-brane
superpotential in terms of a period-like integral. 
There goes quite a bit more into this construction, and
the precise details of the geometric setup can be found in 
\cite{Aganagic:2000gs,Aganagic:2001nx,Lerche:2002yw}.

Actually it is not just expressions for the superpotential on
both sides what gets mapped onto each other by mirror symmetry, but also the
open string modulus $\hat t$, and more generally there is a whole
sequence of various different chain integrals (``semi-periods'')
of the form (\ref{eq:chainint}) which on the Type IIA correspond to
\beq
\label{eq:semiperiods}
\hat \Pi^{\hat \al}(t,\hat t)\ =\ \big\{\hat t_k,\cW^\ell(t,\hat t),...\big\}\ .
\eeq
They form the open string analog of the closed string ``bulk'' period vector
$\Pi^\al(t)$ (\ref{eq:periodInt}) of section 2.1.

\subsection{Relative homology}

Let us summarize what we found: in the Type IIB picture,
both flux-induced and brane-induced 
superpotentials (\ref{eq:WBfluxpot},\ref{eq:chainint})
have a very similar form,
\begin{eqnarray}
\label{eq:FBpots}
\cW_{{\rm flux}}\ &=&\ N_\al \Pi^\al\ \equiv\ N_\al\int_{\ga3_\al}\Omega^{3,0},\\
\cW_{{\rm brane}}\ &=&\ \hat N_{\hat \a} \hat \Pi^{\hat \a}\ \equiv\ \hat N_{\hat \a}\int_{\gah3_{\hat \a}}\Omega^{3,0}.
\end{eqnarray}
This suggests formulating a framework in which the two sectors are uniformly
treated, by succinctly writing:
\beq
\label{eq:totalp}
\cW_{{\rm total}}\ =\ {\mathcal N}_A \Pi^A\ \equiv\ {\mathcal N}_A
\int_{\Ga3_A}\Omega^{3,0}\ ,
\eeq
where
$$
\Ga3_A\ \in\ \big\{\ga3_\al,\gah3_{\hat \a}\big\}\ \cong\ H_3(\whx,Y,\ZZ)
$$
is the set of {\it relative homology cycles}, defined to be the set
of SL 3-cycles in $\whx$ that need to be closed only up to a
boundary lying on the 2-cycle $\ga2$ that is wrapped by the $D$-brane.
By definition, they belong to the indicated relative homology group,
with $Y\equiv \ga2$.

Correspondingly, the possible superpotentials are given by integral
linear combinations of the components of the ``relative period
vector'':
\begin{eqnarray}
\label{eq:relperiods}
\Pi^A\ \equiv\ (\Pi^\al,\hat \Pi^{\hat \a})\ &=&\ (1,t_I,\cW^J,....),\ {\rm where}\\
t_I\ &=&\ \{t_i,\hat t_k\}\nonumber\\
\cW^J\ &=&\ \{\del^i\cF(t),\cW^\ell(t,\hat t)\}\ ,\nonumber
\end{eqnarray}
which one may call the ``holomorphic potentials of $N=1$ Special
Geometry". This emphasizes that they are the analogs of the prepotential
$\cF$, which is governed by $N=2$ Special Geometry. Note that a
subset of them, namely those which come from the closed 3-cycles,
are not independent but are derivatives of the bulk prepotential
$\cF$. This reflects that the flux-induced superpotentials arise
from the spontaneous breaking of $N=2$ supersymmetry to $N=1$, and
that they do not carry more information than what the bulk, closed
string geometry already provides. On the other hand, the brane-induced
potentials $\cW^\ell$ are more genuine $N=1$ quantities, and this
is reflected by the fact that in general they do not integrate to
a generating function; the index $\ell$ labels the $D$-branes, and
there is one value for each possible boundary component. The much
larger variety of the $\cW^\ell$ reflects the obvious fact that
$N=1$ supersymmetric lagrangians are less restricted than the $N=2$
supersymmetric ones.

At this point one may ask what insight can be gained by apparently
just combining the flux and brane sectors into a larger set, and
attaching the label ``relative'' on it.  The point is, and this is
a priori not at all obvious, that closed and open string sectors
consistently fit together into one larger moduli space. As we will
point out below, the total moduli space is flat, despite not being
a direct product of closed and open string moduli spaces; in fact,
the open string sector is deformed by the closed one (but not vice
versa), and it may be viewed as a fibration over the closed one.

In this larger, flat moduli space, we can for example study
singularities that appear when appropriately adjusting both open
and closed string moduli, and investigate the monodromies that are
induced by encircling such singularities. Generic monodromies will
mix the components of the relative period vector (\ref{eq:relperiods}),
and thus mix brane with flux numbers.\foot{Even without further
analysis, we can tell beforehand that such monodromies will always
modify brane numbers by adding or subtracting flux numbers, but not vice versa: this
follows from the fact that there is always a monodromic ambiguity
in adding a closed 3-cycle to an open one, but not the other way
around. This is one manifestation of the stated fact that brane
quantities are deformed by bulk quantities, but not
vice versa (at least in the TFT at tree level, 
which is insensitive to back-reaction).

Note also that this brane-flux mixing is different to the large-$N$
brane-flux transitions of ref.~\cite{Vafa:2000wi},
 which can make branes disappear
and turn into flux configurations; taking brane or flux numbers to
be large is nowhere important in our discussion.} This kind of
quantum duality symmetries, which generalize the well-known
$Sp(2h^{2,1}+2,\ZZ)$-valued monodromy transformations of the bulk
theory, is an example of an insight that can be gained by a uniform
treatment of flux and brane induced potentials.

\subsection{The Special Geometry underlying the $N=1$ superpotential}

Just like the $N=2$ prepotential, the $N=1$ superpotential
(\ref{eq:totalp}) (given by periods and semi-periods) can be
interpreted from at least three interrelated viewpoints \cite{Lerche:2002ck}:

\begin{itemize} 
\item{\bf A)} As part of the $d=4$ $N=1$ supersymmetric space-time
{\bf effective lagrangian}. From that perspective, the flux and
brane induced superpotentials have a special structure as compared
to an arbitrary effective superpotential: as discussed above, they have an
integral instanton expansion of the form $\sum d_{n,\hat
n}{\rm Li}_2[q^n\hat q^{\hat n}]$ (the flux induced ones being a
small subset of this), and they have specific transformation
properties under duality transformations (they are sections
over the combined open/closed string moduli space).

\item{\bf B)} Generating function of {\bf open string TFT correlators}
on the disk.  The main difference to what we discussed before is that
we now deal with a CFT on world-sheets with boundaries, and the
structure of the observables and their correlators are correspondingly
modified. In particular, in the topological $B$-model we consider
$B$-type (Dirichlet) boundary conditions along the sub-manifold
$Y\equiv\ga2$, around which the $D$-brane is wrapped:
$$
\psi^i\ =\ 0\ \ (D)\ ,\qquad\ \lambda_i\ =\ 0\ \ (N)\ ,
$$
where $\lambda,\psi$ are the left- and right-moving fermions of the
non-linear sigma model on the Calabi-Yau threefold $\whx$.  In terms
of these fields, the boundary chiral ring observables $\hat O^{p,q}$
formally look like as in Section 2.2, but are now interpreted as
elements of\foot{While this is true for the specific geometries we consider,
more generally the boundary ring is given by an appropriate Ext group;
see e.g., \cite{Sharpe:2003dr}.} $H^{0,q}(Y,\wedge^pN_Y)$ rather than of
$H^{0,q}(Y,\wedge^pT(\whx))$; here $N_Y$ denotes the normal
bundle to the 2-cycle $Y$. The generators of the boundary ring are
fermionic\foot{That the boundary chiral ring has fermionic generators
has first been emphasized in ref.~\cite{PMMcKay}.  Note that the
deformations of the theory are nevertheless described by {\it
bosonic} parameters $\hat t$, because the 1-form operators will be
integrated against a supercharge; i.e., the perturbing boundary
terms in the 2d lagrangian look like $\hat t_k\int_x G^-\hat
O^{1}_k$.} because they are associated with 1-forms:
\beq
\label{eq:bringgen}
\hat O^{1}_{\hat a}\ =\ \omega^{1(\hat a)}_i\lambda_i\ \in H^0(Y,N_Y)\ .
\eeq
These open string moduli operators generate the boundary chiral ring as follows,
and moreover obey mixed bulk-boundary operator products as indicated:
\begin{eqnarray}
\label{eq:bring}
\cR^{(\partial)}:\qquad\hat O^1_{{\hat a}}\cdot\hat O^1_{{\hat b}}\
&=&\ \sum_{\hat c} C_{{\hat a}{\hat b}}^{\partial\ {\hat c}}O^2_{{\hat c}}\ ,\\
O^{-1,1}_{B,{a}}\cdot\hat O^1_{{\hat b}}\
&=&\ \sum_{\hat c} C_{{a}{\hat b}}^{B\, {\hat c}}O^2_{B,{\hat c}}\ ,\nonumber
\end{eqnarray}
while the bulk (anti-chiral, chiral) ring remains as before. All
of these OPE's have an interpretation in terms of cup products of
certain cohomology groups, and this is explained in more detail in 
ref.~\cite{Lerche:2002yw}. Suffice it to observe from the OPE structure that
the boundary ring is deformed by the bulk deformations, but not
vice versa - this is another manifestation of the above-mentioned
non-renormalization property.

The upshot is that we can formally pull through the program
of $N=2$ Special Geometry, but now for {\it relative cohomology}
instead of the absolute one. To facilitate this, we can implement the
extension of the bulk chiral ring by boundary operators by introducing
formal superfields 
\beq
\label{eq:relcoho}
\vec\cO_A^1\ =\ (O^{-1,1}_\al,\hat O^1_{\hat \al})\ \in H^*(\whx,Y)\ ,
\eeq
that are elements of a relative cohomology group --
by definition the one which is dual to 
the relative homology group of section~3.4. 
Hence the OPE's shown above can concisely be written as:
\beq
\label{eq:relring}
\cR^{open/closed}:\qquad\vec\cO_A^1\cdot \vec\cO_B^1\ =\ 
\sum_C C_{AB}^{\ C}\,\vec\cO_C^2\ .
\eeq
Of course, if there are more boundary components than one,
there will be correspondingly more components of the ``superfields''.

The concept of relative cohomology is very natural in
the context of $D$-branes, and we will see momentarily that it
provides the right framework for doing explicit calculations,
such as deriving the analogs of the Picard-Fuchs differential
equations in the presence of $D$-branes.

\item{\bf C)} Viewpoint of {\bf Hodge variations} in relative cohomology.
Note that the notation (\ref{eq:relcoho}) precisely mirrors the
structure of differentials in relative cohomology: for $Y$ being a
sub-manifold of $X$, these generically look like
$$
\vec\Theta\ =\ (\theta_X,\theta_Y),\ \ \theta_X\in H^*(X),\ \theta_Y\in H^*(Y)\ ,
$$
and their equivalence relation (modulo addition of exact forms) is:
\beq
\label{eq:equivrel}
\vec\Theta\ \cong\ \vec\Theta+(d\omega,i^*\omega-d\eta)\ .
\eeq
Here, $\omega$ and $\eta$ are forms of one degree less than $\theta_X$
and $\theta_Y$, the $i^*$ is the pullback of $\omega$ onto $Y$.
This relation tells that if some differential form is exact on
$X$ and thus trivial in $H^*(X)$, it may actually be non-trivial on
a sub-manifold $Y$. We can understand this also by loosely saying
that total derivatives can become non-trivial under integrals once
there are boundaries, ie.: $\int_\gamma
d\lambda=\int_{\partial\gamma}\lambda$. Translated into physics
terms, this means that operators that are BRST-exact in the closed
string theory, can become non-trivial when $D$-branes are present.

The natural pairing between relative homology and cohomology is then
given by
\beq
\label{eq:pairing}
\langle\,\Gamma_A,\,\vec\Theta_B\,\rangle
\ =\ \int_{\Gamma_A}\theta^{(\whx)}_\be-
\int_{\partial\Gamma_A}\theta^{(Y)}_{\hat \be}\ 
\equiv\ \Pi^A_B(z(t),\hat z(t,\hat t))\ ,
\eeq
which is invariant under the equivalence relation (\ref{eq:equivrel}). By definition,
this yields the relative period matrix, which in the
present context looks as follows:
\beq
\label{eq:relperiodmatrix}
\Pi^A_B(t,\hat t)\ =\ \left(\begin{array}{cccc}
1\ &\ \{t_i,\hat t_k\}\ &\ \{\partial_a{\mathcal F},\, {\mathcal W}^\ell\}\ &\ ... \\
0\ &\ 1\ &\ \partial_I\{\partial_a{\mathcal F},\, {\mathcal W}^\ell\}\  & ... \\
0 & ... & ... & ...\\
\end{array} \right)\ ,
\eeq
and whose top row $\Pi^A\equiv \Pi^A_1$ is the relative period
vector containing the flux and brane superpotentials.  The dots to
the right indicate further unspecified terms, which would appear
in a general situation but which do not appear for the non-compact
geometries we consider.

We have seen in Section 2.2 how the variation of Hodge structures
in the B-model geometry leads to a system of differential equations
for the period matrix, which in turn allows to compute the prepotential
$\cF$ explicitly. In the present context of $N=1$ Special Geometry,
we deal in addition with semi-periods based on chain integrals, and
thus the question is whether we can generalize the method of variation of Hodge
structures in order to find differential equations which would
determine the relative period matrix (\ref{eq:relperiodmatrix}).

In fact we have already prepared the ground for doing
so, and all what is necessary is to follow the same logic as before
with regard to Hodge variations, but now for relative cohomology.
This will automatically take care in a systematic way of boundary terms that arise from
the sub-manifold $Y$. Explicitly, the way it
works is easiest illustrated by the following diagram:
$$
\xymatrix{
\vec\Omega\equiv (\Omthree30\whx,0)\ar[r]^{\sst \delta_z}\ar[dr]^{\sst \delh}&
(\Omthree21\whx,0)\ar[r]^{\sst \delta_z}\ar[dr]^{\sst \delh}&
(\Omthree12\whx,0)\ar[r]^{\sst \delta_z}\ar[dr]^{\sst \delh}&
(\Omthree03\whx,0)\ar[dr]^{\sst \delta_z,\delh} \\
&(0,\Omthree20Y)\ar[r]^{\sst \delta_z,\delh}&
(0,\Omthree11Y)\ar[r]^{\sst \delta_z,\delh}&
(0,\Omthree02Y)\ar[r]^{\sst \delta_z,\delh}& 0}
$$
Here, $\delta_z$ denotes variations with respect to the complex
structure moduli of the threefold $\whx$, and enter in the top row
precisely as discussed before (these arrows correspond to the
rightmost terms in eq.~(\ref{eq:Hodgevar}), the other terms are
not shown).  The new ingredient are the variations with respect to
the brane moduli $\hat z$. These open up another branch in the
diagram, ie., the bottom row, which is associated with differential forms
localized on the sub-manifold
$Y\equiv \partial\gah3$. Evidently this diagram applies to a
situation with only one boundary component, or $D$-brane; for several
components $Y_i$, there will be a branch for every $i$. Moreover,
for intersecting $Y_i$ there can be further sub-branches localized on
the intersections, and so on.

Precisely as explained earlier, one can translate this diagram into a system
of matrix differential equations of the form:
\begin{eqnarray}
\nabla_a\cdot \Pi^A_B\ &=\ \big(\partial_{z_a}-\cA_a(z)\big)\cdot
\Pi^A_B(z,\hat z)\ =&\ 0\\
\nabla_{\hat a}\cdot \Pi^A_B\ &=\ \big(\partial_{\hat z_{\hat a}}-
\cA_{\hat a}(z,\hat z)\big)\cdot\Pi^A_B(z,\hat z)\ =&\ 0\ ,\nonumber
\end{eqnarray}
which represents the open string extension of the PF system.\foot{The
derivation and structure of such extended PF systems has been
discussed in \cite{Mayr:2001xk,Govindarajan:2001zk,Lerche:2001cw,Lerche:2002yw};
the original derivation \cite{Mayr:2001xk} was done
from the viewpoint of Calabi-Yau fourfolds, whose BPS geometry is
dual to the presently discussed Calabi-Yau threefolds with wrapped
$D$-branes.} Again, by recursive elimination of the lower rows
of $\Pi^A_B$, one can rewrite it in terms of higher order differential
operators acting on the top row, i.e., the relative period vector.
In order to help the reader, we will present an explicit example
in the next section.

However, before doing so, let us emphasize again that what we are
not just simply tensoring
the open and closed string sectors. Rather, the open and closed
string moduli form a combined moduli space which is not a direct
product, and it is thus highly non-trivial that it is flat (see the
detailed discussion in \cite{Lerche:2002yw}). This means that:
$$
\big[\nabla_a,\nabla_b\big]\ =\ \big[\nabla_a,\nabla_{\hat b}\big]\ =\
\big[\nabla_{\hat a},\nabla_{\hat b}\big]\ =\ 0\ ,
$$
and this allows to go to flat coordinates, which
 are just the K\"ahler-type open and closed string moduli
$\hat t_k,t_i$ of the $A$-model. For these flat coordinates the
Gau\ss-Manin connection vanishes, so that covariant derivatives
become ordinary ones and in particular, we can write for the open/closed
chiral ring structure constants (\ref{eq:bring}):
\beq
\label{eq:cderivW}
C_{IJ}^{\ \ell}(t,\hat t)\ =\ \partial_I\partial_J\cW^\ell(t,\hat t)\ .
\eeq
This is the open string analog of the well-known property
(\ref{eq:cderivF}) of the bulk chiral ring structure constants,
which integrate to the prepotential $\cF(t)$. However, as we already
have said, in the open string sector the $\cW^\ell$ are in general
not derivatives of a single generating function $\cF$, which reflects
the greater freedom of $N=1$ supersymmetric theories; rather, the
index $\ell$ labels the independent boundary, or $D$-brane sectors.

\end{itemize}

\subsection{An explicit example}

To illustrate our ideas and demonstrate that things
work as claimed, we now work out a concrete example.
Specifically we will consider a non-compact Calabi-Yau threefold
$X$ given by the canonical bundle on $\IP^2$. In simple terms,
it is given by taking the $\IP^2$ (which is not a Calabi-Yau manifold
due to its non-vanishing first Chern class), and adding an extra non-compact
dimension to it such that the first Chern class is cancelled and
the whole manifold is Calabi-Yau. This happens if the extra coordinate
has the correct degree (or charge, in linear sigma model language)
equal to $-3$, whence one usually denotes the resulting threefold
by $X=O(-3)_{\IP^2}$. 

This non-compact Calabi-Yau manifold
 and its mirror geometry has been thoroughly discussed
in the literature \cite{Hori:2000kt,HIV,Aganagic:2001nx}, 
and specifically the
superpotential for a $D6$-brane in this geometry has been computed
before \cite{Aganagic:2001nx}. It was also used as prime example in our work 
\cite{Lerche:2002yw} that
we report about here; however, that discussion was based on the
toric geometry of $O(-3)_{\IP^2}$, and this is 
beyond the scope of this lecture. Rather, we will present here
an alternative (and somewhat over-simplified) calculation which does not
rely on toric geometry, and invite the interested reader to consult
ref.~\cite{Lerche:2002yw} for a more elegant and efficient derivation based
on toric geometry (as well as for discussion of many subtleties 
and details that we drop here, such as the framing ambiguity).

As is well-known \cite{Hori:2000kt,HIV}, the mirror geometry $\whx$ of the 
non-compact Calabi-Yau manifold $O(-3)_{\IP^2}$
is characterized by a LG theory with superpotential,
\beq
\label{eq:Ptwopot}
W(y_i,\zo)\ =\  y_0+y_1+y_2+\zo {y_0^3\over y_1 y_2}\ ,
\eeq
where $\zo\sim e^{-t_1}$ is the complex structure modulus.
It is convenient to write the period integrals in the form
\beq
\label{eq:perint}
\int_{\ga3}\Omega^{3,0}(\zo)\ =\ \int_{\ga3}{1\over P(\zo)}d\mu\ ,
\eeq
where
$$
P = y_0 (x_1^2 + x_2^2) + y_1y_2 W(y_i,z)\ ,
$$ 
and where the measure is:
$$
d\mu  =  {dy_0 dy_1 dy_2 \over y_1 y_2} dx_1 dx_2 \ .
$$
From this we see that $y_{1,2}$ are $\IC^*$, or exponential, variables.

As explained in \cite{Aganagic:2000gs,Aganagic:2001nx},
the $D$-brane geometry can be specified by
imposing extra linear relations between the coordinates $y_i$. 
In fact there are several inequivalent possibilities (``phases'') for
doing so, which lead to different superpotentials. We will consider
in the following 
the boundary condition $y_0=-\hat zy_1$, which corresponds to
the ``outer phase'' of ref.~\cite{Aganagic:2001nx}. Actually we will perform
most computations in terms of the more convenient variable
$\xi=-1/\hat z$, and transform back to $\hat z$ at the end.
Moreover we will first rescale
$y_1\to \zx y_1$, so that we have the simpler boundary condition
\beq
\label{eq: Ptwobc}
H:\qquad y_1\ =\ y_0\ .
\eeq
By this rescaling, the dependence on the open string modulus
$\hat z$ has been transferred from the boundary condition to
the holomorphic 3-form $\Omega^{3,0}$, (\ref{eq:perint}). 
Given this $\Omega^{3,0}(z,\hat z)$,
we now choose a basis of differentials as follows:
\begin{eqnarray}
\label{eq:formbasis}
\vec\vpi
&=&
\Big(\,
\Omega, \,
\delta_\zo \Omega,\,
d\eta, \,
{(\delta_\zo)^2 \Omega},\,
\delta_\zo d\eta\,
\Big)
\\
&= &
\Big(
{1\over P} d\mu, 
-{{y_0}^3\over P^2} d\mu,
{1\over\zx}y_1\partial_1\Big[ {1\over P}\Big]d\mu, 
{2{y_0}^6\over P^3},
-{1\over\zx}y_1\partial_1\Big[ {{y_0}^3\over P^2}\Big]d\mu
\Big)\nonumber
\end{eqnarray}
where $\partial_i\equiv {\partial\over \partial y_i}$ and
$d\eta\equiv\delta_\zx\Omega$ is a form that is exact on $\whx$.
However, under chain integrals with non-zero boundary we have
$\int_\gamma d\eta=\int_{\partial\gamma}\eta$, reflecting that the exact
form $d\eta$ is a non-zero element in the relative cohomology,  localized
on the hyperplane $H$:
\begin{eqnarray}
\label{eq:hyperpl}
d\eta\ =\ {1\over\zx}y_1\partial_1\Big[ {1\over P}\Big]d\mu 
\ \to\ \ \ \ \ \ \eta\ &=&\ -{1\over\zx}{1\over P_\bdy}d\mu_\bdy
\equiv -{1\over\zx}{1\over W_\bdy}d\nu_\bdy,
\\
\delta_\zo d\eta\ =\,-{1\over\zx}y_1\partial_1\Big[ {{y_0}^3\over P^2}\Big]d\mu
\ \to\ \ \ \delta_\zo \eta\ &=&\  {1\over\zx}{{y_0}^3\over {P_\bdy}^2}d\mu_\bdy
\equiv {1\over\zx} {{y_0}^2\over {W_\bdy}^2}d\nu_\bdy.\nonumber
\end{eqnarray}
From this we see that on $H$, the $(2,0)$ form $\eta$ can
formally associated with a ``boundary potential'', given by 
\begin{eqnarray}
\label{eq:wbdy}
P_\bdy\ &=&\  P\vert_{y_1=y_0}\ =\ y_0 W_\bdy
\\
W_\bdy &=& \zx(1+\zx) y_0 y_2 +  \zx y_2^2 + \zo y_0^2+x_1^2 + x_2^2\ ,
\nonumber
\end{eqnarray}
and with the measure:
\beq
d\mu_\bdy  =  {dy_0  dy_2 \over  y_2} dx_1 dx_2 \ .
\eeq
As indicated in (\ref{eq:hyperpl}), one may alternatively consider
$W_\del$ as a boundary potential and instead use the following
measure:
\beq
d\nu_\bdy  =  {dy_0  dy_2 \over y_0 y_2} dx_1 dx_2\ .
\eeq
Note that here $y_0$ has been turned from a $\IC$ to a $\IC^*$ variable.

One can then expand, as usual,
derivatives of the differentials $\vec\vpi$
back into differentials modulo exact pieces, 
however keeping track of those exact pieces that 
possibly contribute under chain integrals.
From partial integration follows:
\beq
{\sum_i p_i(y)\partial_i  P(y)\over  P(y)^{\ell+1}}
=
{1\over \ell}\Big( 
{\sum_i\partial_i p_i(y) \over  P(y)^{\ell}}
-
\sum_i\partial_i \Big[{p_i(y) \over  P(y)^{\ell}}\Big]
\Big)
\eeq
For the chain integrals with boundary condition $y_1 = y_0$, 
the term to the right gives additional contributions:
\beq
\sum_i\partial_i \Big[{p_i(y) \over  P(y)^{\ell}}\Big]=
{({1\over y_0} p_0(y)-p_1(y))\vert_{y_1 = y_0}
\over
P_\bdy(y)^\ell}
\eeq
The numerator can then be further expanded
into the boundary differentials
$(\eta,\delta_\zo\eta)$ plus vanishing relations of the form 
${\partial i} P_\bdy(y)$; that is, the boundary 
sector behaves like a
autonomous LG theory with potential $P_\bdy(y)$. Equivalently, we 
can write everything in terms of the reduced boundary potential $W_\bdy(y)$,
provided we use logarithmic derivatives also with respect to $y_0$.

Eventually, by iteratively applying this procedure and collecting all the terms, 
one can write a matrix representation of the $\zo,\zx$-derivatives 
acting on the basis (\ref{eq:formbasis}) of forms. Transforming back to the
coordinate $\zn=-1/\zx$, this can be written as
\begin{eqnarray}
\label{eq:MatSys}
\nabla_A\, \vec\vpi \ &=& \ 0\ , \qquad A=0,1\ ,
\\
(\nabla_A)_i^{\ j}\ &=&\ 
\delta_i^{\ j}{\partial\over\partial z_A}-(\cA_A)_i^{\ j}\ ,
\nonumber
\end{eqnarray}
where $A$ labels the open and closed string sectors 
($z_0\equiv\hat z$, $z_1\equiv z$) and where
\begin{eqnarray}
\label{eq:Amatr}
\cA_0&=&
\begin{pmatrix}     
0 & 0 & 1 & 0 & 0 \\ 0 & 0 & 0 & 0 & 1 \\ 0 & 0 & 
  -{\frac{1}{\zn}} & 0 & {\frac{\left( -3 + \zn \right) \,\zo}
    {\left( -1 + \zn \right) \,\zn}} \\ 0 & 0 & 0 & 0 & 
  -{\frac{1 - 2\,\zn + {{\zn}^2} + 6\,{{\zn}^3}\,\zo}
     {\zo\,\left( 1 - 2\,\zn + {{\zn}^2} + 4\,{{\zn}^3}\,\zo \right) }}
   \\ 0 & 0 & 0 & 0 & {\frac{1 - 3\,\zn + 3\,{{\zn}^2} - 
      6\,{{\zn}^4}\,\zo + {{\zn}^3}\,\left( -1 + 10\,\zo \right) }{
      \left( -1 + \zn \right) \,\zn\,
      \left( 1 - 2\,\zn + {{\zn}^2} + 4\,{{\zn}^3}\,\zo \right) }} \\  
\end{pmatrix}      
\\
\cA_1&=&
\begin{pmatrix}    
 0 & 1 & 0 & 0 & 0 \\ 0 & 0 & 0 & 1 & 0 \\ 0 & 0 & 0
   & 0 & 1 \\ 0 & {\frac{-1 - 60\,\zo}
    {{{\zo}^2}\,\left( 1 + 27\,\zo \right) }} & 
  {\frac{2\,\zn}{{{\zo}^2} + 27\,{{\zo}^3}}} & 
  {\frac{-3\,\left( 1 + 36\,\zo \right) }{\zo + 27\,{{\zo}^2}}} & 
  {\frac{\zn\,\left( -9 + 31\,\zn - 37\,{{\zn}^2} + 
        {{\zn}^3}\,\left( 17 - 18\,\zo \right)  + 
        {{\zn}^4}\,\left( -2 + 22\,\zo \right)  \right) }{\left( -1 + 
        \zn \right) \,\zo\,\left( 1 + 27\,\zo \right) \,
      \left( 1 - 2\,\zn + {{\zn}^2} + 4\,{{\zn}^3}\,\zo \right) }} \\ 0
   & 0 & 0 & 0 & -{\frac{1 - 2\,\zn + {{\zn}^2} + 6\,{{\zn}^3}\,\zo}
     {\zo\,\left( 1 - 2\,\zn + {{\zn}^2} + 4\,{{\zn}^3}\,\zo \right) }}
   \end{pmatrix} \ .  \nonumber  
\end{eqnarray}
It is straightforward to verify:
\beq
\label{eq:}
\big[\nabla_A\,,\,\nabla_B\big]\ =\ 0\ ,
\eeq
which expresses the advertised flatness and integrability of the combined 
closed and open string moduli space.

Upon iterative elimination of the higher components of $\vec\omega$,
one can reduce the first-order system (\ref{eq:MatSys}) to a system
of Picard-Fuchs operators acting on $\int\Omega^{3,0}$:
\begin{eqnarray}
\label{eq:Lops}
{\cL}_1&=&
{{\xt(1)}^2}\left( \xt(1) - \xt(0) \right)  +
  \zo\left( 3\xt(1) - \xt(0) \right)
   \left( 1 + 3\xt(1) - \xt(0) \right)
   \left( 2 + 3\xt(1) - \xt(0) \right)\ ,
\\
{\cL}_0&=&
\left( 3\xt(1) - \xt(0) \right) \xt(0)-
\zn\left( \xt(1) - \xt(0) \right) \xt(0)\nonumber\ ,
\end{eqnarray}
where $\theta_A\equiv z_A{\del\over\del_{z_A}}$.
This coincides with the PF system derived in \cite{Mayr:2001xk,Lerche:2002ck}. 
The solutions of this system were discussed in \cite{Lerche:2001cw}
and in particular include the mirror maps:
\begin{eqnarray}
\label{eq:flatcoPtwoIII}
\xo(\zo)\ &=&\  \log(\zo)-3 A(\zo)\ ,\\
\xs(\zo,\zn)\ &=&\ \log(\zn)+A(\zo)\ ,\nonumber\\
A(\zo)\ &=&\ -\sum_{{n>0}}
{(-)^{n}(3n-1)!\over (n!)^3}
 {\zo}^{n}\ .\nonumber
\end{eqnarray}
Note that the closed string coordinate $t$ does not depend on
the open string variable $\hat z$, however the open one depends on both the open
and closed ones. This reflects what we said before, namely that the
part of the moduli space pertaining to the open string sector is
deformed by the closed string sector, but not vice versa; in other words we
have a fibration of the open over the closed string moduli space.

One can check that when transforming to the flat coordinates,
the matrices $\cA_A(z,\hat z)$ indeed turn into upper-triangular matrices 
$C_A(\xo,\xs)$,
which amounts to a vanishing Gau\ss-Manin connection, $\Gamma_A$
(the vanishing of the diagonal terms also requires certain rescalings of the basis of differential forms, see below). 
The linear system (\ref{eq:MatSys}) then takes the form: 
$$
\Big({\partial\over \partial t_I}\ -\ C_I(t)\,\Big)\cdot 
\Pi^A_B(t,\hat t)\ = 0\ .
\qquad I=0,1\ .
$$
Its solutions form the columns of the relative period
matrix $\Pi^A_B$, defined by integrating a basis of differential
forms $\omega_i$ over a suitable set of 3-cycles and chains.
Concretely, we choose the variations of $\Omega^{3,0}$ with respect
to the flat coordinates as new basis for the differential forms (in
particular, we define $d\eta=\delta_\xs\Omega^{3,0}$).  Then,
defining the cycles $\ga3_\alpha$ (for $\alpha=1,2,4$) and chains
$\gah3_\alpha$ (for $\alpha=3,5$) such that the first row $\Pi^A_1$
yields the known solutions of the PF system {\ref{eq:Lops}), we can
write the enlarged period matrix as follows:
\beq
\label{eq:RelPerM}
\Pi^A_B(\xo,\xs)\ =\
{\scriptstyle
\begin{pmatrix}
\int_{\ga3_1}\!\!\Omega &
\int_{\ga3_2}\!\!\Omega &
\int_{\gah3_3}\!\!\Omega &
\int_{\ga3_4}\!\!\Omega &
\int_{\gah3_5}\!\!\Omega
\\
0 &
\int_{\ga3_2}\!\!\delta_{\xo}\Omega &
0 &
\int_{\ga3_4}\!\!\delta_{\xo}\Omega &
\int_{\gah3_5}\!\!\delta_{\xo}\Omega
\\
0 &
0 &
\int_{\partial\gah3_3}\!\!\eta &
0 &
\int_{\partial\gah3_5}\!\!\eta
\\
0 &
0 &
0 &
f\int_{\ga3_4}\!\!(\delta_{\xo})^2\Omega &
f\int_{\gah3_5}\!\!(\delta_{\xo})^2\Omega
\\
0 &
0 &
0 &
0 &
g\int_{\partial\gah3_5}\!\!\delta_{\xo}\eta
\end{pmatrix}
}
\eeq
where
$f=1/\del_{\xo}^3\cF(t)$  and $g=1/\del_{\xs}\del_{\xo}\cW(\xo,\xs)$ 
are the rescalings necessary to achieve a completely vanishing connection.
In this basis (up to a further minor rotation), 
the chiral ring structure constants look in terms of the flat coordinates:
\begin{eqnarray}
\label{eq:Cmatrices}
C_0(\xo,\xs)&=&
\begin{pmatrix} 
0 & 0 & 1 & 0 & 0 \\ 
0 & 0 & 0 & 0 & \del_{\xs}\del_{\xo}\cW\\ 
0 & 0 & 0 & 0 &  \del_{\xs}^2\cW \\
0 & 0 & 0 & 0 &  0 \\ 
0 & 0 & 0 & 0 & 0 \\  
\end{pmatrix}
\\
C_1(\xo,\xs)&=&
\begin{pmatrix}
 0 & 1 & 0 & 0 & 0 \\
0 & 0 & 0 & \del_{\xo}^3\cF & \del_{\xo}^2\cW \\
0 & 0 & 0 & 0 & \del_{\xs}\del_{\xo}\cW \\
0 & 0 & 0 & 0 & 0\\
0 & 0 & 0 & 0 & 0 \\ 
\end{pmatrix}
\ ,
\end{eqnarray}
while the relative period matrix turns into:
\beq
\Pi^A_B(\xo,\xs)\ = \ 
\begin{pmatrix} 
1 &\xo & \xs &\del_{\xo}\cF & \cW
\\
0 &1  & 0 &\del_{\xo}^2\cF & \del_{\xo}\cW
\\
0 &0 & 1 & 0  & \del_{\xs}\cW
\\
0 &0  & 0 & 1 & 0
\\
0 &0  & 0 &0 & 1
\\
\end{pmatrix}
\eeq
The top row displays the components of the relative period vector,
which are the building blocks for the total effective, flux and
brane induced superpotential.

Above, $\cF=\cF(\xo)$ indeed coincides with the bulk $N=2$ prepotential
associated with $O[-3]_{\IP^2}$ (which is of the form: $\del_{\xo}^3\cF=
-{1\over3}+\sum n^3d_n{q^n\over 1-q^n}$), and $\cW=\cW(\xo,\xs)$
with the superpotential on the world-volume of the $D6$-brane.
Note, however, that the superpotential obtained in this way:
\begin{eqnarray}
\label{eq:finalpot}
\cW(\xo,\xs)\ &=&\ \frac1{12}(\xo+3\xs)^2 + \cW_0(\zo(\xo),\zn(\xo,\xs)) \ ,
\\
\cW_0(\zo,\zn)\ &=&\ \sum_{{n\geq0,\hat n>n}}
{\frac{{{\left( - \right) }^{n}}
     \left( \hat n - n -1\right) !}{\left( \hat n -
        3n \right) !{(n!)^2}\hat n}}
{\zn}^{{\hat n}} {\zo}^{n}
\end{eqnarray}
has a non-vanishing classical term. This reflects a subtlety 
\cite{Lerche:2002yw} in
that the geometry we were studying is actually not precisely the
one we claimed, and to correct for this one needs to subtract the
classical term in (\ref{eq:finalpot}). The non-perturbative 
piece indeed has integral expansion coefficients $d$ when written
in terms of the flat coordinates: 
$$
\cW_0(\zo(\xo),\zn(\xs,\xo))\ =\ 
\sum_{n,\hat n}d_{n,\hat n}{\rm Li}_2(e^{-nt-\hat n\hat t})\ ,
$$
whose explicit numerical values can be found in the literature.

The potential (\ref{eq:finalpot}) is a generalized hypergeometric
function, and can be analytically continued over the whole open/closed
string moduli space.  At each point one can go to suitable flat
coordinates and study the physics in a local neighborhood, for
example near non-perturbative critical points. In the semi-classical
regime where $z,\hat z$ are small, the powers of $\hat z$ are always
larger than the powers of $z$, and this means that the potential
can be minimized by sending $\hat z\rightarrow0$, independently of
what value $z$ has - this means that the brane ``runs off to
infinity'' while the bulk geometry stays fixed.

In contrast, this is not so in the other possible phase 
\cite{Aganagic:2001nx} of the
theory (or patch, where we put the brane). One can compute the
superpotential in a similar way and obtain \cite{Lerche:2001cw}:
\beq
\label{eq:innerW}
\cW_0(\zo,\zn)\ =\ \sum_{{n,\hat n\geq0,n\not=\hat n}}
{\frac{{{\left( - \right) }^{n}}\left( \hat n + 2n -1\right)
!}{\hat n!{(n!)^2} \left( \hat n - n \right) }}{\hat z}^{\hat n} {z}^{n}\ ,
\eeq
and one finds that the bulk modulus $z$ appears not always
multiplied with the brane modulus $\hat z$; this means that the
presence of the brane wants to make $z$ small as well, and thus
drives the Calabi-Yau to the large radius, or decompactification
limit.

\section{Conclusions}

We have shown how the computation of superpotentials of certain $N=1$
supersymmetric string vacua can be put on a footing very similar
to the computation of the prepotential $\cF$ of $N=2$ supersymmetric
theories. Such $N=1$ vacua can be very crudely labelled by $(X,N_a;\hat
N_{\hat a})$, where $X$ is a compact or non-compact Calabi-Yau
threefold and $N_a$, $\hat N_{\hat a}$ denote flux and $D$-brane
numbers. Switching on fluxes involves only the geometry of the
bulk, ie., closed string sector on $X$, which is governed by $N=2$
Special Geometry. On the other hand, putting $D$-branes produces
more genuine $N=1$ theories, without hidden $N=2$ special geometry.

Often geometries with $D$-branes are dual to closed string
backgrounds, or equivalent to them to after a large-$N$ transition,
the main example given by the conifold which is dual, after a
large-$N$ transition, to a pure flux configuration \cite{Vafa:2000wi} on a
non-compact three-fold. However, the conifold is a very special,
degenerate case, and in line of what we were just saying, in general
a $D$-brane configuration is {\it not} dual to some flux configuration
on a threefold - the functional diversity that can appear in the
superpotential is much larger for $D$-branes as compared to flux
vacua, not the least because the branes bring extra moduli into the
game (such as their positions or Wilson lines), while for flux vacua
only the bulk moduli enter. 

Nevertheless one may ask whether there are other types of closed string
backgrounds that are dual to $D$-branes on threefolds. Indeed, it
has been found that the geometry we have been discussing here
(relative cohomology of non-compact toric threefolds), is dual to
the one of non-compact toric {\it fourfolds} with fluxes; this has
been discussed in detail in refs. \cite{Mayr:2001xk,Lerche:2001cw}. 
Thus, when trying to
classify $N=1$ vacua in terms of flux backgrounds, one might better
consider fourfolds rather than threefolds in order to get a more
accurate counting of possibilities.  Flux-induced superpotentials
on fourfolds (given by certain period integrals), 
have been discussed first in 
\cite{Mayr:1996sh,Lerche:1997zb} 
and \cite{Gukov:1999ya}, and subsequently in many other papers.

It should be clear to the reader that we have barely scratched the surface
of the subject; indeed we have kept under the rug all sorts of subtleties and
details, which can be found in the original papers 
\cite{Lerche:2002ck,Lerche:2002yw}; see also \cite{Sharpe:2003dr} for further aspects and clarification from a more mathematical viewpoint.
For example, we have neglected the so-called framing ambiguity
\cite{Aganagic:2001nx},
which is labelled by an integer $\nu$ and which reflects boundary
conditions at infinity for non-compact branes. Explicitly, for the geometry
we have been considering in the previous section, one obtains the following infinite sequence of superpotentials:
\beq
\label{eq:framingW}
\cW^{(\nu)}(\zo,\zn)\ =\ \sum_{n,{\hat n}} {
(-)^{n+{\hat n}\nu}\, \Gamma(-n+(\nu+1)\, {\hat n})\Gamma({\hat n}) \over
\Gamma(-3n+{\hat n}+1)\Gamma(n+\nu {\hat n} +1)\Gamma(n+1)\Gamma({\hat n}+1)}
\zo^{n}\zn^{{\hat n}}\ ,
\eeq
and the computation we have shown just corresponds to $\nu=0$. 
This demonstrates that when we talk about non-compact branes,
we also need to specify the framing $\nu$ besides the data $(X,N_a;\hat
N_{\hat a})$ in order to define the $D$-brane geometry. And this is 
not the complete story, because we can put the $D$-branes also in
different patches (``phases'') of $X$, and this produces even further
possibilities for obtaining different and inequivalent superpotentials.
Moreover, recall also that we have considered only a very specific class of
$D$-brane geometries (in $A$-model language: wrapped 2-cycles without extra gauge bundles).

All in all, we see that the systematic investigation on superpotentials
in string theories with $N=1$ supersymmetry is a problem of enormous
complexity.  However, there is great hope for making progress toward
a deeper understanding of $N=1$ vacua, as well as the quantum
geometry underlying them, and this is especially due to the recent works 
\cite{AKMV1,ADKMV,Iqbal:2003ds}.
 
Let me conclude by presenting below 
a picture which superficially resembles the final picture
of many talks on string unification, however its meaning is totally different --
it does not show the moduli space of some string theory, rather it shows
various {\it approaches} for tackling $N=1$ string vacua.

\begin{figure}[htb]
\epsfysize=3.in
\centerline{\epsffile{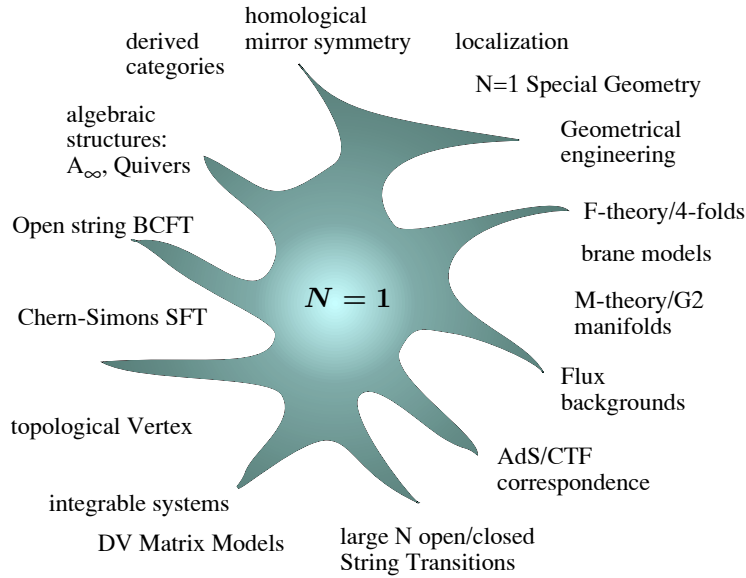}}
\caption{\footnotesize 
There are many viewpoints from which one can tackle string vacua with
$N=1$ supersymmetry, each of which has its own scope, merits and limitations.
What we have covered in these lectures is a small neighborhood of
``$N=1$ Special Geometry''. \label{fig:FIG2}}
 \end{figure}

\section*{Acknowledgments}

I thank the organizers of the School for providing me a pleasant opportunity
to present this material, and Peter Mayr and Nick Warner for a fruitful collaboration
on these matters.


\addcontentsline{toc}{section}{References}

\begingroup\raggedright\endgroup

\end{document}